\documentclass[10pt, twocolumn, notitlepage, superscriptaddress, prb, longbibliography]{revtex4-2}

\usepackage{here}
\usepackage{amsmath}         
\usepackage{graphicx}
\usepackage{braket}         
\usepackage{lettrine}
\usepackage{xcolor}
\usepackage{times}
\usepackage{hhline}
\usepackage{tabularx}
\usepackage{calc}
\usepackage{soul}
\usepackage{mathtools}

\usepackage{amssymb}

\usepackage[colorlinks=true,linkcolor=blue, citecolor=blue, urlcolor=blue]{hyperref}%

\newcolumntype{P}[1]{>{\centering\arraybackslash}p{#1}}
\newcolumntype{M}[1]{>{\centering\arraybackslash}m{#1}}

\begin{document}

\title{
First-principles calculation of orbital Hall effect by Wannier interpolation:
\\
Role of orbital dependence of the anomalous position}

\author{Dongwook Go}
\email{dongo@uni-mainz.de}
\affiliation{Institute of Physics, Johannes Gutenberg University Mainz, 55099 Mainz, Germany}


\author{Hyun-Woo Lee}
\affiliation{Department of Physics, Pohang University of Science and Technology, Pohang 37673, Korea \looseness=-1}

\author{Peter M. Oppeneer}
\affiliation{Department of Physics and Astronomy, Uppsala University, P.O. Box 516, SE-75120 Uppsala, Sweden}

\author{Stefan Bl\"ugel}
\affiliation{Peter Gr\"unberg Institut, Forschungszentrum J\"ulich and JARA, 52425 J\"ulich, Germany \looseness=-1}

\author{Yuriy Mokrousov}
\affiliation{Institute of Physics, Johannes Gutenberg University Mainz, 55099 Mainz, Germany}
\affiliation{Peter Gr\"unberg Institut, Forschungszentrum J\"ulich and JARA, 52425 J\"ulich, Germany \looseness=-1}

\begin{abstract}
The position operator in a Bloch representation acquires a gauge correction in the momentum space on top of the canonical position, which is called the anomalous position. We show that the anomalous position is generally orbital-dependent and thus plays a crucial role in the description of the intrinsic orbital Hall effect in terms of Wannier basis. We demonstrate this from the first-principles calculation of orbital Hall conductivities of transition metals by Wannier interpolation. Our results show that consistent treatment of the velocity operator by including the correction term originating from the anomalous position predicts the orbital Hall conductivities different from those obtained by considering only the group velocity. We find the difference is crucial in several metals. For example, we predict the negative sign of the orbital Hall conductivities for elements in the groups X and XI such as Cu, Ag, Au, and Pd, for which the previous studies predicted the positive sign. Our work suggests the importance of consistently describing the spatial dependence of basis functions by first-principles methods as it is fundamentally missing in the tight-binding approximation. 
\end{abstract}

\date{\today}                 
\maketitle

\section{Introduction}

A realization that orbitally polarized electrons can flow in a solid opened a new paradigm of research on the transport of the orbital degree of freedom of electrons, which is now called \emph{orbitronics}~\cite{Go2021}. Seminal works by Bernevig {\it et al.} and by Kontani \emph{et al.} predicted that orbital current can be electrically induced by orbital Hall effect (OHE) in semiconductors~\cite{Bernevig2005} and transition metals~\cite{Kontani2009}, respectively, by which an external electric field generates transverse flow of electrons depending on the direction of orbital angular momentum (OAM). However, because OAM is \emph{quenched} in equilibrium~\cite{Kittel_textbook}, it was assumed for more than a decade that OHE is stable only in the presence of strong spin-orbit coupling (SOC) and generally suppressed in metals. Thus, theories focused on electrons near the high-symmetry points in the Brillouin zone in semiconductors, where the orbital degeneracy is present~\cite{Bernevig2005} or the OAM is unquenched due to broken inversion symmetry~\cite{Bhowal2020, Canonico2020, Bhowal2021, Cysne2022, Sahu2021}. Nonetheless, some of us have shown that the orbital quenching in equilibrium does not necessarily suppress OHE because it is intrinsically a \emph{non-equilibrium} phenomenon and an external electric field induces the hybridization that results in coherent superposition carrying finite OAM~\cite{Go2018}. Moreover, in the intrinsic mechanism, OHE works as a precursor to spin Hall effect (SHE) $-$ SHE can be understood as a concomitant effect of OHE in the presence of SOC, not the other way around~\cite{Kontani2009, Go2018}. Thus, theories predict that OHE is large even in $3d$ metals, in which SHE is smaller by an order of magnitude~\cite{Jo2018, Salemi2022}.

In recent years, various experimental works have reported the evidence of OHE by optical and transport measurements. Choi \emph{et al.} employed the magneto-optical Kerr effect to detect the orbital accumulation at the surface of a Ti thin film, which is driven by the OHE in the bulk~\cite{Choi2023}. Similarly, by using the same technique, Lyalin \emph{et al.} have also measured the current-induced orbital accumulation resulting from the OHE in thin films of Cr~\cite{Lyalin2023}. In spin-orbitronics, following the theoretical prediction that OAM can induce dynamics of the magnetization by transferring its angular momentum to local moments~\cite{Go2020a, Go2020b}, recent magnetotransport experiments performed on spintronic devices confirmed the role of orbital current in current-induced torques~\cite{Kim2021, Lee2021b, Ding2022} and its interconversion to/from spin current~\cite{Ding2020, Lee2021a, Sala2022, Santos2023}. Orbital current has also been detected in ultrafast timescale by terahertz spectroscopy~\cite{Seifert2023, Wang2023, Xu2023, Kumar2023}. However, we emphasize that due to a twin-like phenomenology, OHE and SHE cannot be clearly separated in general. Thus, the focus of recent works has moved to distinctive features of orbital current such as the relaxation and dephasing lengths~\cite{Go2023, Liao2022, Hayashi2023, Bose2023} and the propagation speed~\cite{Seifert2023}. Nonetheless, experiments on OHE inevitably rely on material-specific predictions based on empirical tight-binding models and first-principles calculations. Due to importance of quantitative prediction of OHE, there have been attempts in different directions for more accurate description of OHE. For example, the role of non-local OAM has been addressed by means of the Berry phase theory~\cite{Pezo2022, Cysne2022}. The influence of the crystal field potential has also been considered for predicting the orbital accumulation driven by OHE, which is responsible for the short lifetime of non-equilibrium orbital-polarized states~\cite{Choi2023}.

In this article, we report a crucial role of the so-called anomalous position~\cite{Adams1959, Karplus1954} in the theoretical description of OHE. We show that the anomalous position emerges in $\mathbf{k}$-space, where $\mathbf{k}$ is the crystal momentum, as a gauge correction to the canonical position and originates from the dipole matrix elements between basis functions in real space. We find the crucial contribution to the anomalous position, which is purely a property of basis states~\cite{Wang2006}, which is present regardless of the microscopic details of the Hamiltonian. For example, this is different from the correction term in the effective description of the electron by the hybridization with the positron in the Dirac equation~\cite{Foldy1950, Vignale2010, Bi2013, Ado2023} and the Yafet term induced by the band hybridizations~\cite{Nozieres1973, Winkler2003, Engel2007}. We show that the anomalous position is orbital-dependent, thus crucially affecting the OHE. We also show that this shares a common microscopic feature with the mechanism of orbital Rashba effect. We demonstrate that the anomalous position significantly influences the intrinsic OHE by first-principles calculations for $3d$, $4d$, and $5d$ transition metals between the group IV and the group XI. We find that the anomalous position affects not only the magnitude but also the sign of the OHE. We predict the negative sign of OHE for some elements in the groups X and XI although previous theoretical works predicted the positive sign for all transition metals~\cite{Tanaka2008, Jo2018, Salemi2022}. We remark that the tight-binding approximation has a fundamental limitation because the information on matrix elements of the position operator is missing unless the spatial dependence of basis states is explicitly stated. This suggests the importance of the microscopic details of the basis set used in the computation of OHE and thus makes the first-principles implementations that consistently describe both core and interstitial regions ideal for the computation of OHE.

The rest of the article is organized as follows. In Sec.~\ref{sec:background}, we provide theoretical background by introducing the definition of the anomalous position, its microscopic origin in the orbital degree of freedom, and its influences on OHE. In Sec.~\ref{sec:methods}, we detail the first-principles methods and implementations used for the computation of OHE. The results for transition metals are presented in Sec.~\ref{sec:results} and we discuss their implications in Sec.~\ref{sec:discussion}. Finally, Sec.~\ref{sec:conclusion} concludes the article with a brief summary.

\section{Theoretical background}
\label{sec:background}

\subsection{Anomalous position}
\label{subsec:anomalous_position}

For a Bloch representation in $\mathbf{k}$-space, it is often taken for granted that the position operator is simply given by $\hat{\mathbf{r}}=i\boldsymbol{\nabla}_\mathbf{k}$ and the velocity operator is given by $\hat{\mathbf{v}}(\mathbf{k}) = (1/i\hbar)[\hat{\mathbf{r}}, \hat{\mathcal{H}}(\mathbf{k})] = (1/\hbar)\boldsymbol{\nabla}_\mathbf{k} \hat{\mathcal{H}}_\mathbf{k}$. However, one can show that an additional term appears in the position operator due to the dipole matrix elements between basis functions, which is known as the anomalous position~\cite{Adams1959, Wang2006}. Let us assume a set of Wannier states $\left\{ \ket{n\mathbf{R}}  \right\}$, where $n$ is the Wannier index and $\mathbf{R}$ is the Bravais lattice vector. In the representation in terms of the Wannier states, the position operator can be decomposed into two parts;
\begin{eqnarray}
\label{eq:position}
\hat{\mathbf{r}}
&=&
\hat{\mathbf{r}}_0
+
\delta \hat{\mathbf{r}}
\end{eqnarray}
where 
\begin{eqnarray}
\hat{\mathbf{r}}_0
=
\sum_{n\mathbf{\mathbf{R}}}
\ket{n\mathbf{R}} \mathbf{R} \bra{n\mathbf{R}}
\end{eqnarray}
describes the position of the center of the unit cell and 
\begin{eqnarray}
\delta \hat{\mathbf{r}}
=
\sum_{n\mathbf{R}}\sum_{n'\mathbf{R}'}
\ket{n\mathbf{R}}
\delta \mathbf{r}_{nn'} (\mathbf{R}'-\mathbf{R})
\bra{n'\mathbf{R}'}
\end{eqnarray}
describes the displacement from the center of the unit cell. The dipole matrix element is defined by
\begin{eqnarray}
\delta \mathbf{r}_{nn'} (\mathbf{R}'-\mathbf{R})
&=&
\bra{n\mathbf{R}} (\hat{\mathbf{r}}-\mathbf{R}) \ket{n'\mathbf{R}'}
\nonumber
\\
&=&
\bra{n\mathbf{0}} \hat{\mathbf{r}} \ket{n'\mathbf{R}'-\mathbf{R}}.
\label{eq:dr_matrix}
\end{eqnarray}
Note that $\delta\hat{\mathbf{r}}$ is invariant under a discrete translation by a Bravais lattice vector. We will show that the first term in Eq.~\eqref{eq:position} can be translated into $i\boldsymbol{\nabla}_\mathbf{k}$ in the Bloch basis and the second term gives rise to a gauge correction to it. 

In order to examine algebraic properties of each term in the position operator, let us consider an arbitrary \emph{bounded} operator in the position space
\begin{eqnarray}
\hat{\mathcal{O}}
=
\sum_{n\mathbf{R}}
\sum_{n'\mathbf{R}'}
\ket{n\mathbf{R}}
\mathcal{O}_{nn'} (\mathbf{R}'-\mathbf{R})
\bra{n'\mathbf{R}'},
\end{eqnarray}
where 
\begin{eqnarray}
\mathcal{O}_{nn'} (\mathbf{R}'-\mathbf{R})
&=&
\bra{n\mathbf{R}} \hat{\mathcal{O}} \ket{n'\mathbf{R}'}
\nonumber
\\
&=&
\bra{n\mathbf{0}} \hat{\mathcal{O}} \ket{n'\mathbf{R}'-\mathbf{R}},
\label{eq:O_matrix}
\end{eqnarray}
and the commutator with the position operator
\begin{eqnarray}
[\hat{\mathbf{r}},\ \hat{\mathcal{O}}]
=
[\hat{\mathbf{r}}_0,\ \hat{\mathcal{O}}]
+
[\delta\hat{\mathbf{r}},\ \hat{\mathcal{O}}].
\label{eq:commutator}
\end{eqnarray}
Note the difference between Eqs.~\eqref{eq:dr_matrix} and \eqref{eq:O_matrix} arising from the unboundedness of the position operator. In the commutator [Eq.~\eqref{eq:commutator}], the first and second terms become
\begin{eqnarray}
[\hat{\mathbf{r}}_0,\ \hat{\mathcal{O}}]
=
-\sum_{n\mathbf{R}}
\sum_{n'\mathbf{R}'}
\ket{n\mathbf{R}}
(\mathbf{R}'-\mathbf{R}) \mathcal{O}_{nn'} (\mathbf{R}'-\mathbf{R})
\bra{n'\mathbf{R}'}
\nonumber
\\
\label{eq:commutator1}
\end{eqnarray}
and
\begin{eqnarray}
\label{eq:commutator2}
[\delta \hat{\mathbf{r}},\ \hat{\mathcal{O}}]
&=&
\sum_{n\mathbf{R}}
\sum_{n'\mathbf{R}'}
\sum_{n''\mathbf{R}''}
\ket{n\mathbf{R}} \bra{n''\mathbf{R}''}
\\
& &
\times
\left[
\delta\mathbf{r}_{nn'}(\mathbf{R}'-\mathbf{R}) \mathcal{O}_{n'n''} (\mathbf{R}''-\mathbf{R}')
\right.
\nonumber
\\
& &
\ \ \ \ \
\left.
-\mathcal{O}_{nn'} (\mathbf{R}'-\mathbf{R}) \delta \mathbf{r}_{n'n''} (\mathbf{R}''-\mathbf{R}')
\right]
,
\nonumber
\end{eqnarray}
respectively. 

In order to find the representation of Eqs.~\eqref{eq:commutator1} and \eqref{eq:commutator2} in $\mathbf{k}$-space, let us define Bloch states from the Wannier states
\begin{eqnarray}
\ket{\psi_{n\mathbf{k}}^\mathrm{(W)}}
=
\frac{1}{\sqrt{N}}
\sum_{\mathbf{R}}
e^{i\mathbf{k}\cdot\mathbf{R}}
\ket{n\mathbf{R}},
\label{eq:Wannier_gauge}
\end{eqnarray}
and the inverse transform which is given by
\begin{eqnarray}
\ket{n\mathbf{R}}
=
\frac{1}{\sqrt{N}}
\sum_\mathbf{k}
e^{-i\mathbf{k}\cdot\mathbf{R}}
\ket{\psi_{n\mathbf{k}}^\mathrm{(W)}},
\label{eq:Wannier_gauge_inverse}
\end{eqnarray}
where $N$ is the number of Bravais lattices. We denote this particular choice of Bloch states by \emph{Wannier gauge}, 
as indicated by the superscript ``(W)''. By plugging Eq.~\eqref{eq:Wannier_gauge_inverse} into Eqs.~\eqref{eq:commutator1} and ~\eqref{eq:commutator2}, we obtain
\begin{eqnarray}
[\hat{\mathbf{r}}_0, \hat{\mathcal{O}}]
=
\sum_{\mathbf{k}}\sum_{nn'}
\ket{\psi_{n\mathbf{k}}^\mathrm{(W)}}
\left[
i\boldsymbol{\nabla}_\mathbf{k} \mathcal{O}_{nn'}^\mathrm{(W)} (\mathbf{k})
\right]
\bra{\psi_{n'\mathbf{k}}^\mathrm{(W)}}
\label{eq:action_r0}
\end{eqnarray}
and 
\begin{widetext}
\begin{eqnarray}
[\delta \hat{\mathbf{r}}, \hat{\mathcal{O}}]
&=&
\sum_\mathbf{k}
\sum_{nn'n''}
\ket{\psi_{n\mathbf{k}}^\mathrm{(W)}}
\left[ 
\boldsymbol{\mathcal{A}}_{nn'}^\mathrm{(W)} (\mathbf{k}) \mathcal{O}_{n'n''}^\mathrm{(W)} (\mathbf{k})
-
\mathcal{O}_{nn'}^\mathrm{(W)} (\mathbf{k}) \boldsymbol{\mathcal{A}}_{n'n''}^\mathrm{(W)} (\mathbf{k})
\right] 
\bra{\psi_{n''\mathbf{k}}^\mathrm{(W)}}
,
\label{eq:action_dr}
\end{eqnarray}
\end{widetext}
where 
\begin{eqnarray}
\mathcal{O}_{nn'}^\mathrm{(W)} (\mathbf{k})
=
\sum_\mathbf{R} e^{i\mathbf{k}\cdot\mathbf{R}} \mathcal{O}_{nn'} (\mathbf{R})
\end{eqnarray}
and 
\begin{eqnarray}
\boldsymbol{\mathcal{A}}_{nn'}^\mathrm{(W)} (\mathbf{k})
=
\sum_\mathbf{R} e^{i\mathbf{k}\cdot\mathbf{R}}
\bra{n \mathbf{0}}
\hat{\mathbf{r}}
\ket{n'\mathbf{R}}.
\label{eq:anomalous_position_Wannier}
\end{eqnarray}
We denote $\boldsymbol{\mathcal{A}}^\mathrm{(W)}(\mathbf{k})$ the anomalous position (in the Wannier gauge) throughout this article. Note that this definition is consistent with the definition of the Berry connection~\cite{Berry1984} in the Wannier gauge,
\begin{eqnarray}
\label{eq:Berry_connection_Wannier}
\boldsymbol{\mathcal{A}}_{nn'}^\mathrm{(W)} (\mathbf{k})
=
i
\braket{
u_{n\mathbf{k}}^\mathrm{(W)} 
|
\boldsymbol{\nabla}_\mathbf{k} u_{n'\mathbf{k}}^\mathrm{(W)} 
},
\end{eqnarray}
where $\ket{u_{n\mathbf{k}}^\mathrm{(W)}}=e^{-i\mathbf{k}\cdot\mathbf{r}}\ket{\psi_{n\mathbf{k}}^\mathrm{(W)}}$ is the periodic part of the Bloch state in the Wannier gauge.

Meanwhile, from the actions of $\hat{\mathbf{r}}_0$ [Eq.~\eqref{eq:action_r0}] and $\delta\hat{\mathbf{r}}$ [Eq.~\eqref{eq:action_dr}], we may define the position operator in the Bloch representation (Wannier gauge) in $\mathbf{k}$-space as
\begin{eqnarray}
{\mathbf{r}}_{nn'}^\mathrm{(W)} (\mathbf{k})
=
i \delta_{nn'} \boldsymbol{\nabla_\mathbf{k}} + \boldsymbol{\mathcal{A}}_{nn'}^\mathrm{(W)} (\mathbf{k}).
\label{eq:position_final}
\end{eqnarray}
The velocity operator, which is defined as\begin{eqnarray}
\label{eq:velocity_definition}
\hat{\boldsymbol{v}}
\coloneqq
\frac{1}{i\hbar}
\left[ \hat{\mathbf{r}}, \ \hat{\mathcal{H}} \right],
\end{eqnarray}
is given by
\begin{widetext}
\begin{eqnarray}
\label{eq:velocity_final}
& &
\hat{\boldsymbol{v}}
=
\sum_{\mathbf{k}} \sum_{nn'}
\ket{\psi_{n\mathbf{k}}^\mathrm{(W)}}
\Bigg[ 
\frac{1}{\hbar}
\boldsymbol{\nabla}_\mathbf{k} \mathcal{H}_{nn'}^\mathrm{(W)} (\mathbf{k})
+
\frac{1}{i\hbar}
\sum_{n''}
\left\{
\boldsymbol{\mathcal{A}}_{nn''}^\mathrm{(W)} (\mathbf{k}) \mathcal{H}_{n''n'}^\mathrm{(W)} (\mathbf{k})
-
\mathcal{H}_{nn''}^\mathrm{(W)} (\mathbf{k}) \boldsymbol{\mathcal{A}}_{n''n'}^\mathrm{(W)} (\mathbf{k})
\right\}
\Bigg]
\bra{\psi_{n'\mathbf{k}}^\mathrm{(W)}}
.
\end{eqnarray}
\end{widetext}
Note that the anomalous position gives the correction to the velocity on top of the group-velocity-like term $(1/\hbar)\boldsymbol{\nabla}_\mathbf{k} \hat{\mathcal{H}}_\mathbf{k}$. Equations~\eqref{eq:position_final} and \eqref{eq:velocity_final} constitute the main result of this section.

\subsection{Gauge degree of freedom}
\label{subsec:gauge}

We emphasize that the anomalous position $\boldsymbol{\mathcal{A}}(\mathbf{k})$ depends on the choice of gauge, and thus we specify its gauge as a superscript, e.g. ``(W)'' in Eq.~\eqref{eq:anomalous_position_Wannier}. Let us consider another important gauge, the so-called Hamiltonian gauge that diagonalizes the Hamiltonian 
\begin{eqnarray}
\label{eq:eigenvalue_eq}
\hat{\mathcal{H}} \ket{\psi_{m\mathbf{k}}^\mathrm{(H)}} = E_{m\mathbf{k}} \ket{\psi_{m\mathbf{k}}^\mathrm{(H)}},
\end{eqnarray}
where $m$ and $E_{m\mathbf{k}}$ has physical meanings of the band index and energy eigenvalue in the Hamiltonian gauge. Note the superscript ``(H)'' that specifies the Hamiltonian gauge. We define the unitary matrix that relates the Wannier gauge and the Hamiltonian gauge, whose elements are defined by
\begin{eqnarray}
U_{nm}(\mathbf{k}) =
\braket{
\psi_{n\mathbf{k}}^\mathrm{(W)}
|
\psi_{m\mathbf{k}}^\mathrm{(H)}
}
\label{eq:gauge_transform1}
\end{eqnarray}
such that 
\begin{eqnarray}
\ket{\psi_{m\mathbf{k}}^\mathrm{(H)}}
=
\sum_{n} \ket{\psi_{n\mathbf{k}}^\mathrm{(W)}} U_{nm} (\mathbf{k}).
\label{eq:gauge_transform2}
\end{eqnarray}
Under the gauge transformation, the position operator in Eq.~\eqref{eq:position_final} becomes
\begin{eqnarray}
\label{eq:position_Hamiltonian_gauge}
\mathbf{r}^\mathrm{(H)}_{mm'} (\mathbf{k})
=
i\delta_{mm'} \boldsymbol{\nabla_\mathbf{k}} + \boldsymbol{\mathcal{A}}_{mm'}^\mathrm{(H)} (\mathbf{k})
\end{eqnarray}
where 
\begin{eqnarray}
\boldsymbol{\mathcal{A}}_{mm'}^\mathrm{(H)} (\mathbf{k})
&=&
\sum_{nn'}
[U^\dagger (\mathbf{k})]_{mn} \boldsymbol{\mathcal{A}}_{nn'}^\mathrm{(W)} (\mathbf{k}) U_{n'm'} (\mathbf{k})
\nonumber
\\
& &
+
i
\sum_n [U^\dagger (\mathbf{k})]_{mn} 
[
\boldsymbol{\nabla}_\mathbf{k} U_{nm'} (\mathbf{k})
]
\label{eq:anomalous_position_Hamiltonian_gauge}
\end{eqnarray}
is the anomalous position in the Hamiltonian gauge. Note that the gauge transformation is consistent with the relation between the Berry connection in the Hamiltonian gauge
\begin{eqnarray}
\boldsymbol{\mathcal{A}}_{nn'}^\mathrm{(H)} (\mathbf{k})
=
i
\braket{
u_{n\mathbf{k}}^\mathrm{(H)} 
|
\boldsymbol{\nabla}_\mathbf{k} u_{n'\mathbf{k}}^\mathrm{(H)} 
}
\end{eqnarray}
for the periodic part of the Bloch state in the Hamiltonian gauge $\ket{u_{n\mathbf{k}}^\mathrm{(H)}}=e^{-i\mathbf{k}\cdot\mathbf{r}}\ket{\psi_{n\mathbf{k}}^\mathrm{(H)}}$ and the Berry connection in the Wannier gauge [Eq.~\eqref{eq:Berry_connection_Wannier}]~\cite{Berry1984}.

The second term in Eq.~\eqref{eq:anomalous_position_Hamiltonian_gauge} is given by 
\begin{eqnarray}
& &
\mathcal{D}_{mm'} (\mathbf{k})
\coloneqq 
\sum_n
\left[U^\dagger (\mathbf{k}) \right]_{mn}
\left[ 
\boldsymbol{\nabla}_\mathbf{k} U_{nm'} (\mathbf{k})
\right]
\nonumber
\\
& &
=
\frac{
\sum_{nn'}
\left[ U^\dagger (\mathbf{k}) \right]_{mn}
\left[
\boldsymbol{\nabla}_\mathbf{k} \mathcal{H}_{nn'}^\mathrm{(W)} (\mathbf{k})
\right]
U_{n'm'} (\mathbf{k})
}
{E_{m'\mathbf{k}} - E_{m\mathbf{k}}}.
\label{eq:UdU_term}
\end{eqnarray}
for $m\neq m'$, which can be obtained in the following way. We start from Eqs.~(\ref{eq:eigenvalue_eq}$-$\ref{eq:gauge_transform2}), or equivalently,
\begin{eqnarray}
\sum_{n'}
\mathcal{H}^\mathrm{(W)}_{nn'} (\mathbf{k}) U_{n'm'} (\mathbf{k})
=
E_{m'\mathbf{k}} U_{nm'} (\mathbf{k}).
\label{eq:eigen_U}
\end{eqnarray}
Differentiation with respect to $\mathbf{k}$ leads to
\begin{eqnarray}
& &
\sum_{n'}
\left[
\boldsymbol{\nabla}_\mathbf{k} \mathcal{H}_{nn'}^\mathrm{(W)} (\mathbf{k})
\right]
U_{n'm'} (\mathbf{k})
+
\mathcal{H}_{nn'}^\mathrm{(W)} (\mathbf{k})
\left[
\boldsymbol{\nabla}_\mathbf{k}
U_{n'm'} (\mathbf{k})
\right]
\nonumber 
\\
& &
=
\left[ 
\boldsymbol{\nabla}_\mathbf{k} E_{m'\mathbf{k}}
\right]
U_{nm'} (\mathbf{k})
+
E_{m'\mathbf{k}} 
\left[ 
\boldsymbol{\nabla}_\mathbf{k} U_{nm'} (\mathbf{k})
\right].
\end{eqnarray}
Multiplying by $[U^\dagger (\mathbf{k})]_{mn}$ and summing over index $n$ gives
\begin{eqnarray}
& &
\sum_{nn'}
\left[ U^\dagger (\mathbf{k}) \right]_{mn}
\left[
\boldsymbol{\nabla}_\mathbf{k} \mathcal{H}_{nn'}^\mathrm{(W)} (\mathbf{k})
\right]
U_{n'm'} (\mathbf{k})
\nonumber 
\\
& &
+
\sum_{nn'}
\left[U^\dagger (\mathbf{k}) \right]_{mn}
\mathcal{H}_{nn'}^\mathrm{(W)} (\mathbf{k})
\left[
\boldsymbol{\nabla}_\mathbf{k}
U_{n'm'} (\mathbf{k})
\right]
\nonumber 
\\
&=& 
\left[ 
\boldsymbol{\nabla}_\mathbf{k} E_{m'\mathbf{k}}
\right]
\sum_n
\left[U^\dagger (\mathbf{k}) \right]_{mn}
U_{nm'} (\mathbf{k})
\nonumber 
\\
& & +
E_{m'\mathbf{k}} 
\sum_n
\left[U^\dagger (\mathbf{k}) \right]_{mn}
\left[ 
\boldsymbol{\nabla}_\mathbf{k} U_{nm'} (\mathbf{k})
\right].
\end{eqnarray}
By using $\sum_n \left[U^\dagger (\mathbf{k}) \right]_{mn} \mathcal{H}_{nn'}^\mathrm{(W)} (\mathbf{k}) = \left[U^\dagger (\mathbf{k}) \right]_{mn'} E_{m\mathbf{k}}$ and $\sum_n\left[U^\dagger (\mathbf{k}) \right]_{mn}U_{nm'} (\mathbf{k}) = \delta_{mm'}$, we obtain
\begin{eqnarray}
& &
\sum_{nn'}
\left[ U^\dagger (\mathbf{k}) \right]_{mn}
\left[
\boldsymbol{\nabla}_\mathbf{k} \mathcal{H}_{nn'}^\mathrm{(W)} (\mathbf{k})
\right]
U_{n'm'} (\mathbf{k}) =
\left[ 
\boldsymbol{\nabla}_\mathbf{k} E_{m'\mathbf{k}}
\right]
\delta_{mm'}
\nonumber
\\
& &
\ \ \ \ \ \ \ \ \ \ 
+
(E_{m'\mathbf{k}} - E_{m\mathbf{k}})
\sum_n
\left[U^\dagger (\mathbf{k}) \right]_{mn}
\left[ 
\boldsymbol{\nabla}_\mathbf{k} U_{nm'} (\mathbf{k})
\right].
\end{eqnarray}
Thus, for $m\neq m'$, $\delta_{mm'}$ disappears and we obtain Eq.~\eqref{eq:UdU_term}.

In the Hamiltonian gauge, the velocity operator [Eq.~\eqref{eq:velocity_final}] becomes
\begin{widetext}
\begin{subequations}
\begin{eqnarray}
\label{eq:velocity_Hamiltonian_gauge}
\hat{\boldsymbol{v}}
&=&
\sum_{\mathbf{k}} \sum_{mm'}
\ket{\psi_{m\mathbf{k}}^\mathrm{(H)}}
\Bigg[ 
\frac{1}{\hbar}
\boldsymbol{\nabla}_\mathbf{k} \mathcal{H}_{mm'}^\mathrm{(H)} (\mathbf{k})
+
\frac{1}{i\hbar}
\sum_{m''}
\left\{
\boldsymbol{\mathcal{A}}_{mm''}^\mathrm{(H)} (\mathbf{k}) \mathcal{H}_{m''m'}^\mathrm{(H)} (\mathbf{k})
-
\mathcal{H}_{mm''}^\mathrm{(H)} (\mathbf{k}) \boldsymbol{\mathcal{A}}_{m''m'}^\mathrm{(H)} (\mathbf{k})
\right\}
\Bigg]
\bra{\psi_{m'\mathbf{k}}^\mathrm{(H)}}
\\
&=&
\sum_{\mathbf{k}} \sum_{mm'}
\ket{\psi_{m\mathbf{k}}^\mathrm{(H)}}
\Bigg[ 
\frac{1}{\hbar}
\delta_{mm'}
\boldsymbol{\nabla}_\mathbf{k} E_{m\mathbf{k}}
+
\frac{1}{i\hbar}
\left(
E_{m'\mathbf{k}} - E_{m\mathbf{k}}
\right)
\boldsymbol{\mathcal{A}}_{mm'}^\mathrm{(H)} (\mathbf{k}) 
\Bigg]
\bra{\psi_{m'\mathbf{k}}^\mathrm{(H)}}
.
\end{eqnarray}
\end{subequations}
\end{widetext}
where in the second line, we have used $\mathcal{H}_{mm'}^\mathrm{(H)}(\mathbf{k}) = \delta_{mm'} E_{m\mathbf{k}}$ in the Hamiltonian gauge. By comparing Eqs.~\eqref{eq:position_final} and Eqs.~\eqref{eq:velocity_final} with Eqs.~\eqref{eq:position_Hamiltonian_gauge} and \eqref{eq:velocity_Hamiltonian_gauge}, respectively, one find that the position and velocity operators are \emph{gauge covariant}. Note that the anomalous position is crucial for the covariance. By Eqs.~\eqref{eq:anomalous_position_Hamiltonian_gauge} and \eqref{eq:UdU_term}, the velocity operator in the Hamiltonian gauge can be computed by the expression in the Wannier gauge and the unitary transformation $U(\mathbf{k})$,
\begin{eqnarray}
\boldsymbol{v}_{mm'}^\mathrm{(H)}
&=&
\frac{1}{\hbar}
\delta_{mm'}
\boldsymbol{\nabla}_\mathbf{k} E_{m\mathbf{k}}
\end{eqnarray}
if $m = m'$, and 
\begin{eqnarray}
\label{eq:velocity_Hamiltonian_gauge_U}
\boldsymbol{v}_{mm'}^\mathrm{(H)}
&=&
\frac{1}{\hbar}
\sum_{nn'}
\left[ U^\dagger (\mathbf{k}) \right]_{mn}
\left[
\boldsymbol{\nabla}_\mathbf{k} \mathcal{H}_{nn'}^\mathrm{(W)} (\mathbf{k})
\right]
U_{n'm'} (\mathbf{k})
\\
& &
+
\frac{1}{i\hbar}
\sum_{nn'}
[U^\dagger (\mathbf{k})]_{mn} 
\left[ 
\boldsymbol{\mathcal{A}}^\mathrm{(W)} (\mathbf{k}), \ 
\mathcal{H}^\mathrm{(W)}
\right]_{nn'}
U_{n'm'} (\mathbf{k})
\nonumber
\end{eqnarray}
if $m\neq m'$. Note the second term in Eq.~\eqref{eq:velocity_Hamiltonian_gauge_U} which appears as a gauge correction to the group-velocity-like term.

\subsection{Orbital origin of the anomalous position}
\label{subsec:orbital_origin_A}

The formal derivation of the anomalous position and gauge-covariant description in the Wannier basis, which we have just discussed in Secs.~\ref{subsec:anomalous_position} and \ref{subsec:gauge}, can also be found in previous works, e.g. in Refs.~\cite{Wang2006, Lopez2012}. However, the microscopic nature of the anomalous position has not been studied so far, to our best knowledge. Here, we show that the anomalous position originates from the orbital degree of freedom of the Wannier functions thus generally affecting the orbital properties strongly. This can be seen already from Eq.~\eqref{eq:dr_matrix}, which is an orbitally-sensitive dipole matrix element between localized Wannier states. 

\begin{figure}[t!]
\centering
\includegraphics[angle=0, width=0.45\textwidth]{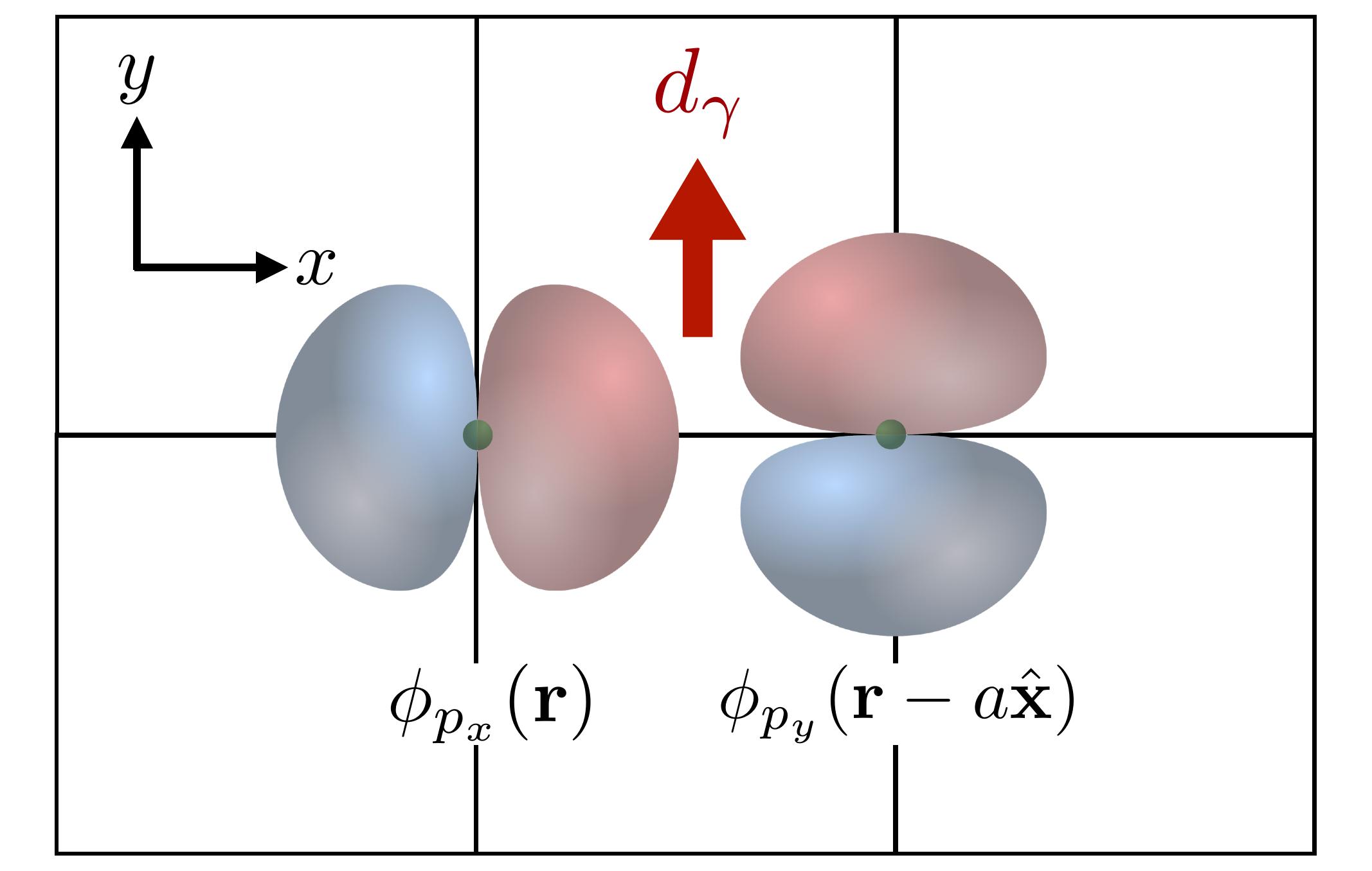}
\caption{
\label{fig:anomalous_position}
Schematic illustration of the matrix element for the anomalous position for $p_x$- and $p_y$-like Wannier states in a two-dimensional cubic lattice, $d_\gamma = \bra{p_x, \mathbf{0}} \hat{y} \ket{p_y, +\hat{\mathbf{x}}a}$ [Eq.~\eqref{eq:d_gamma}].}
\end{figure}

To demonstrate this further, let us consider a two-dimensional lattice model with $p_x$, $p_y$, and $p_z$ Wannier states as an example [Fig.~\ref{fig:anomalous_position}]. If we consider only the nearest neighbor couplings for the dipole matrix elements, Eq.~\eqref{eq:anomalous_position_Wannier} results in the following expression for the anomalous position:
\begin{widetext}
\begin{eqnarray}
{\boldsymbol{\mathcal{A}}}^\mathrm{(W)}(\mathbf{k})
&=&
e^{+ik_x a}
\begin{pmatrix}
\bra{p_x, \mathbf{0}} \hat{\mathbf{r}} \ket{p_x, +\hat{\mathbf{x}}a} & 
\bra{p_x, \mathbf{0}} \hat{\mathbf{r}} \ket{p_y, +\hat{\mathbf{x}}a} &
\bra{p_x, \mathbf{0}} \hat{\mathbf{r}} \ket{p_z, +\hat{\mathbf{x}}a} \\
\bra{p_y, \mathbf{0}} \hat{\mathbf{r}} \ket{p_x, +\hat{\mathbf{x}}a} & 
\bra{p_y, \mathbf{0}} \hat{\mathbf{r}} \ket{p_y, +\hat{\mathbf{x}}a} &
\bra{p_y, \mathbf{0}} \hat{\mathbf{r}} \ket{p_z, +\hat{\mathbf{x}}a} \\
\bra{p_z, \mathbf{0}} \hat{\mathbf{r}} \ket{p_x, +\hat{\mathbf{x}}a} & 
\bra{p_z, \mathbf{0}} \hat{\mathbf{r}} \ket{p_y, +\hat{\mathbf{x}}a} &
\bra{p_z, \mathbf{0}} \hat{\mathbf{r}} \ket{p_z, +\hat{\mathbf{x}}a}
\end{pmatrix}
+
\nonumber
\\
& &
e^{-ik_x a}
\begin{pmatrix}
\bra{p_x, \mathbf{0}} \hat{\mathbf{r}} \ket{p_x, -\hat{\mathbf{x}}a} & 
\bra{p_x, \mathbf{0}} \hat{\mathbf{r}} \ket{p_y, -\hat{\mathbf{x}}a} &
\bra{p_x, \mathbf{0}} \hat{\mathbf{r}} \ket{p_z, -\hat{\mathbf{x}}a} \\
\bra{p_y, \mathbf{0}} \hat{\mathbf{r}} \ket{p_x, -\hat{\mathbf{x}}a} & 
\bra{p_y, \mathbf{0}} \hat{\mathbf{r}} \ket{p_y, -\hat{\mathbf{x}}a} &
\bra{p_y, \mathbf{0}} \hat{\mathbf{r}} \ket{p_z, -\hat{\mathbf{x}}a} \\
\bra{p_z, \mathbf{0}} \hat{\mathbf{r}} \ket{p_x, -\hat{\mathbf{x}}a} & 
\bra{p_z, \mathbf{0}} \hat{\mathbf{r}} \ket{p_y, -\hat{\mathbf{x}}a} &
\bra{p_z, \mathbf{0}} \hat{\mathbf{r}} \ket{p_z, -\hat{\mathbf{x}}a}
\end{pmatrix}
+
\nonumber
\\
& &
e^{+ik_y a}
\begin{pmatrix}
\bra{p_x, \mathbf{0}} \hat{\mathbf{r}} \ket{p_x, +\hat{\mathbf{y}}a} & 
\bra{p_x, \mathbf{0}} \hat{\mathbf{r}} \ket{p_y, +\hat{\mathbf{y}}a} &
\bra{p_x, \mathbf{0}} \hat{\mathbf{r}} \ket{p_z, +\hat{\mathbf{y}}a} \\
\bra{p_y, \mathbf{0}} \hat{\mathbf{r}} \ket{p_x, +\hat{\mathbf{y}}a} & 
\bra{p_y, \mathbf{0}} \hat{\mathbf{r}} \ket{p_y, +\hat{\mathbf{y}}a} &
\bra{p_y, \mathbf{0}} \hat{\mathbf{r}} \ket{p_z, +\hat{\mathbf{y}}a} \\
\bra{p_z, \mathbf{0}} \hat{\mathbf{r}} \ket{p_x, +\hat{\mathbf{y}}a} & 
\bra{p_z, \mathbf{0}} \hat{\mathbf{r}} \ket{p_y, +\hat{\mathbf{y}}a} &
\bra{p_z, \mathbf{0}} \hat{\mathbf{r}} \ket{p_z, +\hat{\mathbf{y}}a}
\end{pmatrix}
+
\nonumber
\\
& &
e^{-ik_y a}
\begin{pmatrix}
\bra{p_x, \mathbf{0}} \hat{\mathbf{r}} \ket{p_x, -\hat{\mathbf{y}}a} & 
\bra{p_x, \mathbf{0}} \hat{\mathbf{r}} \ket{p_y, -\hat{\mathbf{y}}a} &
\bra{p_x, \mathbf{0}} \hat{\mathbf{r}} \ket{p_z, -\hat{\mathbf{y}}a} \\
\bra{p_y, \mathbf{0}} \hat{\mathbf{r}} \ket{p_x, -\hat{\mathbf{y}}a} & 
\bra{p_y, \mathbf{0}} \hat{\mathbf{r}} \ket{p_y, -\hat{\mathbf{y}}a} &
\bra{p_y, \mathbf{0}} \hat{\mathbf{r}} \ket{p_z, -\hat{\mathbf{y}}a} \\
\bra{p_z, \mathbf{0}} \hat{\mathbf{r}} \ket{p_x, -\hat{\mathbf{y}}a} & 
\bra{p_z, \mathbf{0}} \hat{\mathbf{r}} \ket{p_y, -\hat{\mathbf{y}}a} &
\bra{p_z, \mathbf{0}} \hat{\mathbf{r}} \ket{p_z, -\hat{\mathbf{y}}a}
\end{pmatrix},
\label{eq:model_1}
\end{eqnarray}
\end{widetext}
where $a$ is the lattice constant. Note that many of the position matrix elements vanish if the integrand is odd with respect to the reflection of either $x$, $y$, or $z$. Thus, the following elements are zero;
\begin{subequations}
\begin{eqnarray}
& & \bra{p_x, \mathbf{0}} \hat{y} \ket{p_x, \pm \hat{\mathbf{x}}a} = 0, \\
& & \bra{p_x, \mathbf{0}} \hat{z} \ket{p_x, \pm \hat{\mathbf{x}}a} = 0, \\
& & \bra{p_x, \mathbf{0}} \hat{x} \ket{p_x, \pm\hat{\mathbf{y}}a} = 0, \\
& & \bra{p_x, \mathbf{0}} \hat{z} \ket{p_x, \pm\hat{\mathbf{y}}a} = 0, 
\label{eq:constraint_parity_1}
\end{eqnarray}
\end{subequations}
\begin{subequations}
\begin{eqnarray}
& & \bra{p_x, \mathbf{0}} \hat{x} \ket{p_y, \pm\hat{\mathbf{x}}a} = 0, \\
& & \bra{p_x, \mathbf{0}} \hat{z} \ket{p_y, \pm\hat{\mathbf{x}}a} = 0, \\
& & \bra{p_x, \mathbf{0}} \hat{y} \ket{p_y, \pm\hat{\mathbf{y}}a} = 0, \\
& & \bra{p_x, \mathbf{0}} \hat{z} \ket{p_y, \pm\hat{\mathbf{y}}a} = 0, 
\label{eq:constraint_parity_2}
\end{eqnarray}
\end{subequations}
\begin{subequations}
\begin{eqnarray}
& & \bra{p_x, \mathbf{0}} \hat{x} \ket{p_z, \pm\hat{\mathbf{x}}a} = 0, \\
& & \bra{p_x, \mathbf{0}} \hat{y} \ket{p_z, \pm\hat{\mathbf{x}}a} = 0, \\
& & \bra{p_x, \mathbf{0}} \hat{x} \ket{p_z, \pm\hat{\mathbf{y}}a} = 0, \\
& & \bra{p_x, \mathbf{0}} \hat{y} \ket{p_z, \pm\hat{\mathbf{y}}a} = 0, \\
& & \bra{p_x, \mathbf{0}} \hat{z} \ket{p_z, \pm\hat{\mathbf{y}}a} = 0, 
\label{eq:constraint_parity_3}
\end{eqnarray}
\end{subequations}
\begin{subequations}
\begin{eqnarray}
& & \bra{p_y, \mathbf{0}} \hat{x} \ket{p_x, \pm\hat{\mathbf{x}}a} = 0, \\
& & \bra{p_y, \mathbf{0}} \hat{z} \ket{p_x, \pm\hat{\mathbf{x}}a} = 0, \\
& & \bra{p_y, \mathbf{0}} \hat{y} \ket{p_x, \pm\hat{\mathbf{y}}a} = 0, \\
& & \bra{p_y, \mathbf{0}} \hat{z} \ket{p_x, \pm\hat{\mathbf{y}}a} = 0, 
\label{eq:constraint_parity_4}
\end{eqnarray}
\end{subequations}
\begin{subequations}
\begin{eqnarray}
& & \bra{p_y, \mathbf{0}} \hat{y} \ket{p_y, \pm\hat{\mathbf{x}}a} = 0, \\
& & \bra{p_y, \mathbf{0}} \hat{z} \ket{p_y, \pm\hat{\mathbf{x}}a} = 0, \\
& & \bra{p_y, \mathbf{0}} \hat{x} \ket{p_y, \pm\hat{\mathbf{y}}a} = 0, \\
& & \bra{p_y, \mathbf{0}} \hat{z} \ket{p_y, \pm\hat{\mathbf{y}}a} = 0, 
\label{eq:constraint_parity_5}
\end{eqnarray}
\end{subequations}
\begin{subequations}
\begin{eqnarray}
& & \bra{p_y, \mathbf{0}} \hat{x} \ket{p_z, \pm\hat{\mathbf{x}}a} = 0, \\
& & \bra{p_y, \mathbf{0}} \hat{y} \ket{p_z, \pm\hat{\mathbf{x}}a} = 0, \\
& & \bra{p_y, \mathbf{0}} \hat{z} \ket{p_z, \pm\hat{\mathbf{x}}a} = 0, \\
& & \bra{p_y, \mathbf{0}} \hat{x} \ket{p_z, \pm\hat{\mathbf{y}}a} = 0, \\
& & \bra{p_y, \mathbf{0}} \hat{y} \ket{p_z, \pm\hat{\mathbf{y}}a} = 0, 
\label{eq:constraint_parity_6}
\end{eqnarray}
\end{subequations}
\begin{subequations}
\begin{eqnarray}
& & \bra{p_z, \mathbf{0}} \hat{x} \ket{p_x, \pm\hat{\mathbf{x}}a} = 0, \\
& & \bra{p_z, \mathbf{0}} \hat{y} \ket{p_x, \pm\hat{\mathbf{x}}a} = 0, \\
& & \bra{p_z, \mathbf{0}} \hat{x} \ket{p_x, \pm\hat{\mathbf{y}}a} = 0, \\
& & \bra{p_z, \mathbf{0}} \hat{y} \ket{p_x, \pm\hat{\mathbf{y}}a} = 0, \\
& & \bra{p_z, \mathbf{0}} \hat{z} \ket{p_x, \pm\hat{\mathbf{y}}a} = 0, 
\label{eq:constraint_parity_7}
\end{eqnarray}
\end{subequations}
\begin{subequations}
\begin{eqnarray}
& & \bra{p_z, \mathbf{0}} \hat{x} \ket{p_y, \pm\hat{\mathbf{x}}a} = 0, \\
& & \bra{p_z, \mathbf{0}} \hat{y} \ket{p_y, \pm\hat{\mathbf{x}}a} = 0, \\
& & \bra{p_z, \mathbf{0}} \hat{z} \ket{p_y, \pm\hat{\mathbf{x}}a} = 0, \\
& & \bra{p_z, \mathbf{0}} \hat{x} \ket{p_y, \pm\hat{\mathbf{y}}a} = 0, \\
& & \bra{p_z, \mathbf{0}} \hat{y} \ket{p_y, \pm\hat{\mathbf{y}}a} = 0, 
\label{eq:constraint_parity_8}
\end{eqnarray}
\end{subequations}
\begin{subequations}
\begin{eqnarray}
& & \bra{p_z, \mathbf{0}} \hat{y} \ket{p_z, \pm\hat{\mathbf{x}}a} = 0, \\
& & \bra{p_z, \mathbf{0}} \hat{z} \ket{p_z, \pm\hat{\mathbf{x}}a} = 0, \\
& & \bra{p_z, \mathbf{0}} \hat{x} \ket{p_z, \pm\hat{\mathbf{y}}a} = 0, \\
& & \bra{p_z, \mathbf{0}} \hat{z} \ket{p_z, \pm\hat{\mathbf{y}}a} = 0.
\label{eq:constraint_parity_9}
\end{eqnarray}
\end{subequations}

Additionally, if the real space representations of Wannier states, e.g. $\phi_{n}(\mathbf{r}-\mathbf{R})\coloneqq \braket{\mathbf{r} | n\mathbf{R}}$, are \emph{real}, where $\bra{\mathbf{r}}$ is the position eigenbra, the position matrix elements satisfy the identity
\begin{eqnarray}
\bra{n\mathbf{0}} \hat{\mathbf{r}} \ket{n'\mathbf{R}}
&=&
\int d^3r \phi_n (\mathbf{r}) \mathbf{r} \phi_{n'} (\mathbf{r}-\mathbf{R})
\nonumber 
\\
&=&
\int d^3r \phi_{n'} (\mathbf{r}-\mathbf{R})  \mathbf{r} \phi_n (\mathbf{r})
\nonumber 
\\
&=&
\int d^3r \phi_{n'} (\mathbf{r})  (\mathbf{r} + \mathbf{R}) \phi_n (\mathbf{r}+\mathbf{R})
\nonumber 
\\
&=&
\int d^3r \phi_{n'} (\mathbf{r})  \mathbf{r}  \phi_n (\mathbf{r}+\mathbf{R})
\nonumber
\\
&=&
\bra{n'\mathbf{0}} \hat{\mathbf{r}} \ket{n-\mathbf{R}},
\label{eq:constraint_real}
\end{eqnarray}
where we have used the orthogonality condition $\mathbf{R}\braket{n'\mathbf{0} | n\mathbf{R}}=0$ on the third line.

By making use of Eqs.~\eqref{eq:constraint_parity_1}-\eqref{eq:constraint_parity_9} and Eq.~\eqref{eq:constraint_real}, Eq.~\eqref{eq:model_1} is simplified to
\begin{widetext}
\begin{eqnarray}
{\boldsymbol{\mathcal{A}}}^\mathrm{(W)}(\mathbf{k})
&=&
e^{+ik_x a}
\begin{pmatrix}
\hat{\mathbf{x}} \bra{p_x, \mathbf{0}} \hat{x} \ket{p_x, +\hat{\mathbf{x}}a} & 
\hat{\mathbf{y}} \bra{p_x, \mathbf{0}} \hat{y} \ket{p_y, +\hat{\mathbf{x}}a} &
\hat{\mathbf{z}} \bra{p_x, \mathbf{0}} \hat{z} \ket{p_z, +\hat{\mathbf{x}}a}
\\
\hat{\mathbf{y}} \bra{p_y, \mathbf{0}} \hat{y} \ket{p_x, +\hat{\mathbf{x}}a} & 
\hat{\mathbf{x}} \bra{p_y, \mathbf{0}} \hat{x} \ket{p_y, +\hat{\mathbf{x}}a} &
0 \\
\hat{\mathbf{z}} \bra{p_z, \mathbf{0}} \hat{z} \ket{p_x, +\hat{\mathbf{x}}a} &
0 &
\hat{\mathbf{x}} \bra{p_z, \mathbf{0}} \hat{x} \ket{p_z, +\hat{\mathbf{x}}a}
\end{pmatrix}
+
\nonumber
\\
& &
e^{-ik_x a}
\begin{pmatrix}
\hat{\mathbf{x}} \bra{p_x, \mathbf{0}} \hat{x} \ket{p_x, +\hat{\mathbf{x}}a} & 
\hat{\mathbf{y}} \bra{p_y, \mathbf{0}} \hat{y} \ket{p_x, +\hat{\mathbf{x}}a} &
\hat{\mathbf{z}} \bra{p_z, \mathbf{0}} \hat{z} \ket{p_x, +\hat{\mathbf{x}}a}
\\
\hat{\mathbf{x}} \bra{p_y, \mathbf{0}} \hat{y} \ket{p_y, +\hat{\mathbf{x}}a} & 
\hat{\mathbf{x}} \bra{p_y, \mathbf{0}} \hat{x} \ket{p_y, +\hat{\mathbf{x}}a} &
0 \\
\hat{\mathbf{z}} \bra{p_x, \mathbf{0}} \hat{z} \ket{p_z, +\hat{\mathbf{x}}a} &
0 &
\hat{\mathbf{x}} \bra{p_z, \mathbf{0}} \hat{x} \ket{p_z, +\hat{\mathbf{x}}a}
\end{pmatrix}
+
\nonumber
\\
& &
e^{+ik_y a}
\begin{pmatrix}
\hat{\mathbf{y}} \bra{p_x, \mathbf{0}} \hat{y} \ket{p_x, +\hat{\mathbf{y}}a} & 
\hat{\mathbf{x}} \bra{p_x, \mathbf{0}} \hat{x} \ket{p_y, +\hat{\mathbf{y}}a} &
0 \\
\hat{\mathbf{x}} \bra{p_y, \mathbf{0}} \hat{x} \ket{p_x, +\hat{\mathbf{y}}a} & 
\hat{\mathbf{y}} \bra{p_y, \mathbf{0}} \hat{y} \ket{p_y, +\hat{\mathbf{y}}a} &
\hat{\mathbf{z}} \bra{p_y, \mathbf{0}} \hat{z} \ket{p_z, +\hat{\mathbf{y}}a} 
\\
0 & 
\hat{\mathbf{z}} \bra{p_z, \mathbf{0}} \hat{z} \ket{p_y, +\hat{\mathbf{y}}a} &
\hat{\mathbf{y}} \bra{p_z, \mathbf{0}} \hat{y} \ket{p_z, +\hat{\mathbf{y}}a} 
\end{pmatrix}
+
\nonumber
\\
& &
e^{-ik_y a}
\begin{pmatrix}
\hat{\mathbf{y}} \bra{p_x, \mathbf{0}} \hat{y} \ket{p_x, +\hat{\mathbf{y}}a} & 
\hat{\mathbf{x}} \bra{p_y, \mathbf{0}} \hat{x} \ket{p_x, +\hat{\mathbf{y}}a} &
0 \\
\hat{\mathbf{x}} \bra{p_x, \mathbf{0}} \hat{x} \ket{p_y, +\hat{\mathbf{y}}a} & 
\hat{\mathbf{y}} \bra{p_y, \mathbf{0}} \hat{y} \ket{p_y, +\hat{\mathbf{y}}a} &
\hat{\mathbf{z}} \bra{p_z, \mathbf{0}} \hat{z} \ket{p_y, +\hat{\mathbf{y}}a} 
\\
0 & 
\hat{\mathbf{z}} \bra{p_y, \mathbf{0}} \hat{z} \ket{p_z, +\hat{\mathbf{y}}a} &
\hat{\mathbf{y}} \bra{p_z, \mathbf{0}} \hat{y} \ket{p_z, +\hat{\mathbf{y}}a} 
\end{pmatrix}.
\label{eq:model_2}
\end{eqnarray}
\end{widetext}
Finally, some of the integrals are identical, so we can define three independent parameters for representing the nonzero dipole matrix elements,
\begin{subequations}
\begin{eqnarray}
d_\sigma 
&\coloneqq & 
- \bra{p_x, \mathbf{0}} \hat{x} \ket{p_x, +\hat{\mathbf{x}}a} 
\\
&=& 
- \bra{p_y, \mathbf{0}} \hat{y} \ket{p_y, +\hat{\mathbf{y}}a},
\end{eqnarray}
\end{subequations}
\begin{subequations}
\begin{eqnarray}
d_\pi 
&\coloneqq &
+ \bra{p_x, \mathbf{0}} \hat{y} \ket{p_x, +\hat{\mathbf{y}}a} 
\\
&=& 
+ \bra{p_y, \mathbf{0}} \hat{x} \ket{p_y, +\hat{\mathbf{x}}a}
\\
&=& 
+ \bra{p_z, \mathbf{0}} \hat{x} \ket{p_z, +\hat{\mathbf{x}}a}
\\
&=& 
+ \bra{p_z, \mathbf{0}} \hat{y} \ket{p_z, +\hat{\mathbf{y}}a},
\end{eqnarray}
\end{subequations}
\begin{subequations}
\begin{eqnarray}
d_\gamma 
&\coloneqq &
+ \bra{p_x, \mathbf{0}} \hat{y} \ket{p_y, +\hat{\mathbf{x}}a}
\\
&=&
- \bra{p_y, \mathbf{0}} \hat{y} \ket{p_x, +\hat{\mathbf{x}}a}
\\
&=&
- \bra{p_x, \mathbf{0}} \hat{x} \ket{p_y, +\hat{\mathbf{y}}a}
\\
&=&
+ \bra{p_y, \mathbf{0}} \hat{x} \ket{p_x, +\hat{\mathbf{y}}a}
\\
&=&
+ \bra{p_x, \mathbf{0}} \hat{z} \ket{p_z, +\hat{\mathbf{x}}a}
\\
&=&
- \bra{p_z, \mathbf{0}} \hat{z} \ket{p_x, +\hat{\mathbf{x}}a}
\\
&=&
+ \bra{p_y, \mathbf{0}} \hat{z} \ket{p_z, +\hat{\mathbf{y}}a}
\\
&=&
- \bra{p_z, \mathbf{0}} \hat{z} \ket{p_y, +\hat{\mathbf{y}}a}.
\label{eq:d_gamma}
\end{eqnarray}
\end{subequations}
For example, a schematic illustration of one of the matrix elements for $\gamma_d$ is shown in Fig.~\ref{fig:anomalous_position}.

Therefore, the final expression of the anomalous velocity for the model becomes
\begin{widetext}
\begin{eqnarray}
{\boldsymbol{\mathcal{A}}}^\mathrm{(W)}(\mathbf{k})
&=&2
\begin{pmatrix}
-d_\sigma \hat{\mathbf{x}}\cos (k_x a) + d_\pi \hat{\mathbf{y}} \cos (k_y a) & 
+id_\gamma \hat{\mathbf{y}} \sin(k_x a) - id_\gamma \hat{\mathbf{x}} \sin (k_y a) &
id_\gamma \hat{\mathbf{z}} \sin (k_x a)
\\
-id_\gamma \hat{\mathbf{y}} \sin(k_x a) + id_\gamma \hat{\mathbf{x}} \sin (k_y a) &
d_\pi \hat{\mathbf{x}}\cos (k_x a) - d_\sigma \hat{\mathbf{y}} \cos (k_y a) & 
id_\gamma \hat{\mathbf{z}} \sin (k_y a)
\\
-id_\gamma \hat{\mathbf{z}} \sin (k_x a) &
-id_\gamma \hat{\mathbf{z}} \sin (k_y a) &
d_\pi \hat{\mathbf{x}}\cos (k_x a) + d_\pi \hat{\mathbf{y}} \cos (k_y a) 
\end{pmatrix},
\nonumber 
\\
\end{eqnarray}
which can be concisely written as
\begin{eqnarray}
{\boldsymbol{\mathcal{A}}}^\mathrm{(W)} (\mathbf{k})
&=&
2\mathrm{diag}
\big[ 
 -\hat{\mathbf{x}} d_\sigma \cos (k_x a) + \hat{\mathbf{y}} d_\pi \cos (k_y a),\ 
+ \hat{\mathbf{x}} d_\pi \cos (k_x a) - \hat{\mathbf{y}} d_\sigma \cos (k_y a),\ 
+ \hat{\mathbf{x}} d_\pi \cos (k_x a) + \hat{\mathbf{y}} d_\pi \cos (k_y a)
\big]
\nonumber 
\\
& & + (2d_\gamma/\hbar)
{L}_z^\mathrm{(W)}
\big[ 
\hat{\mathbf{x}} \sin (k_y a) - \hat{\mathbf{y}} \sin (k_x a)
\big]
+ (2d_\gamma/\hbar)
\hat{\mathbf{z}}
\big[ 
{L}_y^\mathrm{(W)} \sin (k_x a) - {L}_x^\mathrm{(W)} \sin (k_y a)
\big].
\label{eq:anomalous_position_model}
\end{eqnarray}
Here, 
\begin{eqnarray}
L_x^\mathrm{(W)}
\coloneqq 
\hbar
\begin{pmatrix}
0 & 0 & 0 \\
0 & 0 & -i \\ 
0 & +i & 0
\end{pmatrix},
\ \
L_y^\mathrm{(W)}
\coloneqq 
\hbar
\begin{pmatrix}
0 & 0 & +i \\
0 & 0 & 0 \\ 
-i & 0 & 0
\end{pmatrix},
\ \
L_z^\mathrm{(W)}
\coloneqq 
\hbar
\begin{pmatrix}
0 & -i & 0 \\
+i & 0 & 0 \\ 
0 & 0 & 0
\end{pmatrix}
\end{eqnarray}
\end{widetext}
are the matrix representations of the $x$, $y$, $z$ components of the OAM operator in the Wannier gauge, respectively. Equation~\eqref{eq:anomalous_position_model} clearly shows that the anomalous position is orbital-dependent, whose role in the computation of OHE will be explained in Sec.~\ref{sec:results}.


\subsection{Relation to the orbital Rashba coupling}

Reference~\cite{Park2011} proposed a mechanism of the orbital Rashba coupling as the interaction between the electric dipole moment and the potential gradient. Here, we formally show that the orbital Rashba coupling is a manifestation of the anomalous position. In two-dimensional systems like our model, it is important to note that the position operator along $z$ is not simply null (we assume that the film is extended in the $xy$ plane). Although $i\partial_{k_z}$ is undefined due to lack of a discrete translation symmetry along $z$, $\mathcal{A}_z^\mathrm{(W)} (\mathbf{k})$ is well-defined and has a physical meaning of electric dipole moment. Therefore, in the perturbation theory picture in the first order of the structural asymmetry, the orbital Rashba coupling in two-dimensional systems can be written as
\begin{eqnarray}
H_\mathrm{OR}^\mathrm{(W)} (\mathbf{k}) &=& e  \mathcal{E}_z^\mathrm{eff} \mathcal{A}_z^\mathrm{(W)} (\mathbf{k}).
\label{eq:orbital_Rashba_formal}
\end{eqnarray}
Here, $\mathcal{E}_z^\mathrm{eff}$ is the effective electric field characterizing the strength of the inversion symmetry breaking, which include both the potential gradient and asymmetry of the structure.

For our model [Eq.~\eqref{eq:anomalous_position_model}], we have the familiar expression of the orbital Rashba Hamiltonian,
\begin{subequations}
\begin{eqnarray}
H_\mathrm{OR}^\mathrm{(W)} (\mathbf{k}) &=&
\frac{2d_\gamma \mathcal{E}_z^\mathrm{eff}}{\hbar}
\left[ 
{L}_y^\mathrm{(W)} \sin (k_x a) - {L}_x^\mathrm{(W)} \sin (k_y a)
\right]
\nonumber
\\
\\
&\approx &
\frac{\alpha_\mathrm{OR}}{\hbar}
\left[ 
{L}_y^\mathrm{(W)} k_x - {L}_x^\mathrm{(W)} k_y
\right], \ \ \text{near } \mathbf{k}=0.
\nonumber
\\
\end{eqnarray}
\label{eq:orbital_Rashba_model}
\end{subequations}
We have defined the orbital Rashba constant as $\alpha_\mathrm{OR}=2d_\gamma \mathcal{E}_z^\mathrm{eff}a$. We emphasize that representation of the orbital Rashba coupling generally depends on the choice of a gauge. For example, Eq.~\eqref{eq:orbital_Rashba_model} is the expression of the orbital Rashba coupling in the Wannier gauge, which assumes the $p_x$, $p_y$, and $p_z$ Wannier states as basis. However, because the representation of quantum states also varies with the choice of a gauge, measurable effects such as the OAM expectation values and energy splittings are gauge independent.

\section{Computational Methods}
\label{sec:methods}

\subsection{Kubo formula}

We demonstrate the importance of the anomalous position in OHE for real materials by first-principles calculation. We evaluate the following Kubo formula for the orbital Hall conductivity
\begin{eqnarray}
\label{eq:Kubo}
\sigma_\mathrm{OH}
&=&
{e\hbar} \int \frac{d^3k}{(2\pi)^3}
\sum_{mm'}
(f_{m\mathbf{k}} - f_{m'\mathbf{k}})
\\
\nonumber
& &
\times
\mathrm{Im}\left[ 
\frac{
\bra{\psi_{m\mathbf{k}}^\mathrm{(H)}} 
\hat{j}_z^{L_y}
\ket{\psi_{m'\mathbf{k}}^\mathrm{(H)}}
\bra{\psi_{m'\mathbf{k}}^\mathrm{(H)}}
\hat{v}_x
\ket{\psi_{m\mathbf{k}}^\mathrm{(H)}}
}{(E_{m\mathbf{k}} - E_{m'\mathbf{k}})(E_{m\mathbf{k}} - E_{m'\mathbf{k}}+i\Gamma)}
\right],
\end{eqnarray}
where $m$ and $m'$ are band indices in the Hamiltonian gauge, $E_{m\mathbf{k}}$ is its energy eigenvalue of the $m$-th band, and $f_{m\mathbf{k}}$ is the corresponding Fermi-Dirac distribution function ($T=300\ \mathrm{K}$ is assumed for the temperature). Note that $\hat{v}_x$ is the $x$ component of the velocity operator [Eq.~\eqref{eq:velocity_final}], and 
\begin{eqnarray}
\hat{j}_z^{L_y}=\frac{1}{2}\left(  \hat{v}_z \hat{L}_y + \hat{L}_y \hat{v}_z \right)
\label{eq:conventional_orbital_current}
\end{eqnarray}
is the orbital current operator with the velocity in the $z$ direction and the OAM polarized along the $y$ direction. We introduce a phenomenological constant $\Gamma=25\ \mathrm{meV}$ for the convenience of convergence within reasonable number ($\sim 10^7$) of $\mathbf{k}$-points in the integral (see Table~\ref{tab:params}). The spin Hall conductivity $\sigma_\mathrm{SH}$ is computed by the same formula by replacing $\hat{L}_y$ into the $y$ component of the spin operator $\hat{S}_y$ in Eq.~\eqref{eq:conventional_orbital_current}.

In practical implementation, all the operators are represented in the Wannier gauge and transformed into the Hamiltonian gauge. The Bloch states in the Hamiltonian gauge are obtained by diagonalizing the Hamiltonian written in the Wannier gauge, which corresponds to finding a unitary matrix ${U}(\mathbf{k})$ [Eq.~\eqref{eq:eigen_U}]. Then we can evaluate the matrix elements of the orbital current and velocity operators in the Hamiltonian gauge by
\begin{eqnarray}
\bra{\psi_{m\mathbf{k}}^\mathrm{(H)}} 
\hat{j}_z^{L_y}
\ket{\psi_{m'\mathbf{k}}^\mathrm{(H)}}
&=&
\left[ 
{U}^\dagger (\mathbf{k})
\cdot
{j_z^{L_y\mathrm{(W)}}} (\mathbf{k})
\cdot
{U} (\mathbf{k})
\right]_{mm'},
\\
\bra{\psi_{m'\mathbf{k}}^\mathrm{(H)}}
\hat{v}_x
\ket{\psi_{m\mathbf{k}}^\mathrm{(H)}}
&=&
\left[ 
{U}^\dagger (\mathbf{k})
\cdot
{v_{x}^\mathrm{(W)}} (\mathbf{k})
\cdot
{U} (\mathbf{k})
\right]_{m'm},
\end{eqnarray}
as outlined in Eq.~\eqref{eq:velocity_Hamiltonian_gauge_U} for the velocity operator in case of $m\neq m'$. Together with the energy eigenvalues $E_{n\mathbf{k}}$ and $E_{m\mathbf{k}}$, these expressions are plugged into Eq.~\eqref{eq:Kubo}, and the orbital Hall conductivity is obtained.

\begin{figure*}[ht!]
\centering
\includegraphics[angle=0, width=1.0\textwidth]{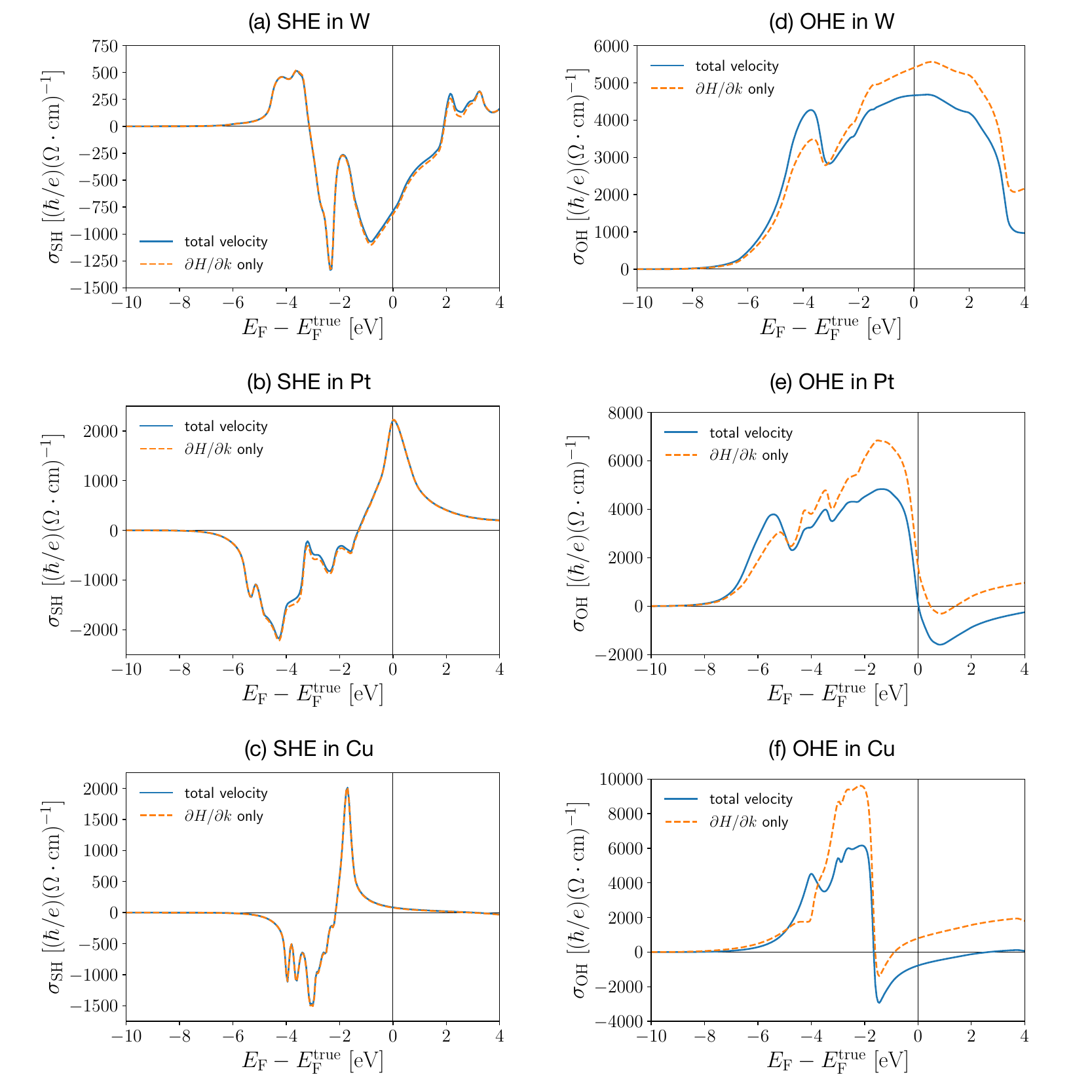}
\caption{
\label{fig:SHE_OHE}
(a,b,c) Intrinsic spin Hall conductivity $\sigma_\mathrm{SH}$ and (d,e,f) orbital Hall conductivity $\sigma_\mathrm{OH}$ of bcc W (a,d; first row), fcc Pt (b,e; second row), and fcc Cu (c,f; third row) as a function of the Fermi energy $E_\mathrm{F}$, which is varied assuming that the effective singe-particle potential is fixed to the potential at the true Fermi energy $E_\mathrm{F}^\mathrm{true}$. Blue solid lines are the results obtained by taking the full expression of the velocity operator in Eq.~\eqref{eq:velocity_final}, and orange dashed lines are the results obtained by taking only the group-velocity-like term and ignoring the anomalous position contribution. 
}
\end{figure*}

\subsection{First-principles calculation}

For describing real materials, we use the \texttt{FLEUR} code~\cite{fleur} that implements the full-potential linearly augmented plane wave (FLAPW) method~\cite{Wimmer1981} of the density functional theory (DFT). We use the Perdew-Burke-Ernzerhof functional within generalized gradient approximation for treating the exchange and correlation effects~\cite{Perdew1996}. The parameters used for the DFT calculation are summarized in Tab.~\ref{tab:params}. Note that the muffin-tin radius $R_\mathrm{MT}$, the plane-wave cutoff $K_\mathrm{max}$, and the maximum of the harmonic expansion in the muffin-tin $l_\mathrm{max}$ approximately satisfy  $R_\mathrm{MT} K_\mathrm{max} \approx l_\mathrm{max}$. This is important for achieving accurate convergence in the FLAPW method which matches the wave function at the muffin-tin boundary~\cite{Singh_LAPW}. We evaluate the OAM operator $\mathbf{L}$ with respect to the center of the atom, which is integrated in the muffin-tin sphere. This is justified for systems where the Bloch states have strong ``atomic'' characters near the atomic centers, which dominantly contribute to the orbital moments. We remark that Hanke {\it et al.} has shown that in bulk transition metals the atom-centered approximation gives the similar value of the orbital moments calculated by means of the Berry phase theory~\cite{Hanke2016}.

In order to construct operators (Hamiltonian, position, spin, and OAM) in the Wannier gauge, we use the \texttt{WANNIER90} code~\cite{Pizzi2020}, which is interface with the \texttt{FLEUR} code~\cite{Freimuth2008}. We use $36 N_\mathrm{atom}$ Kohn-Sham states to construct $18 N_\mathrm{atom}$ maximally localized Wannier states, where $N_\mathrm{atom}$ is the number of atoms in the unit cell. The number 18 comes from the selection of the projections with $s$, $p_x$, $p_y$, $p_z$, $d_{z^2}$, $d_{x^2-y^2}$, $d_{xy}$, $d_{yz}$, $d_{zx}$ shape Wannier states with spin up and down as the initial guess of the Wannier states. We use the frozen energy window in the disentanglement step, whose maximum is set $5\ \mathrm{eV}$ above the Fermi energy. Then we iteratively perform unitary operations to find a set of Wannier states with minimal spread in real space. We remark that the maximally localized Wannier states form a computationally highly efficient basis set, whose number is significantly reduced compared to the number of basis states in the FLAPW description. This is advantageous for evaluating the response function [Eq.~\eqref{eq:Kubo}], which exhibits spiky features in (anti)crossings of bands and thus requires a dense sampling of $\mathbf{k}$-points. The number of points in the interpolation $\mathbf{k}$-mesh is shown in Tab.~\ref{tab:params}, which is significantly denser compared to the DFT $\mathbf{k}$-mesh.

\begin{figure*}[t!]
\centering
\includegraphics[angle=0, width=0.6\textwidth]{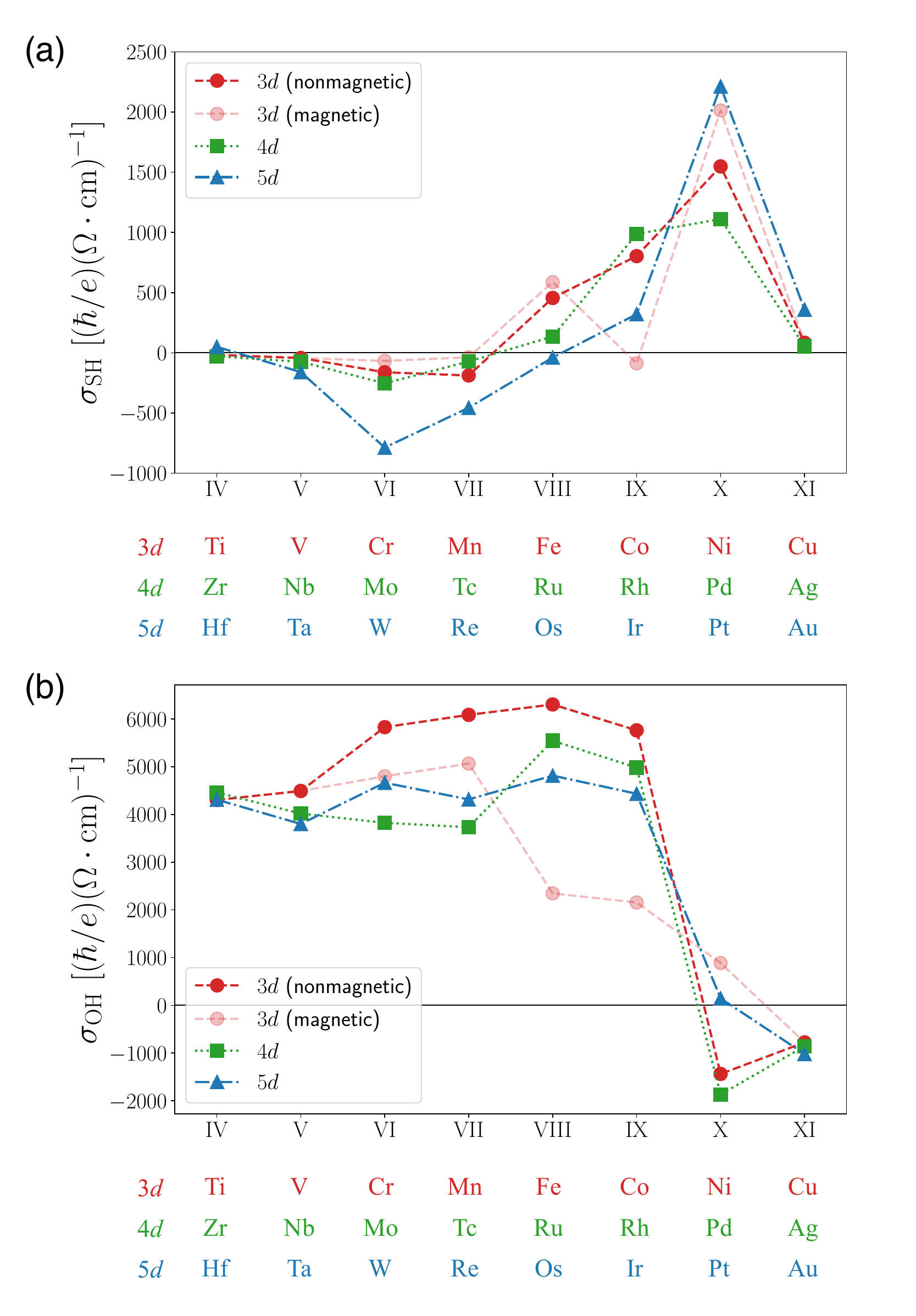}
\caption{
The intrinsic (a) spin Hall conductivity $\sigma_\mathrm{SH}$ and (b) orbital Hall conductivity $\sigma_\mathrm{OH}$ in $3d$ (red circles), $4d$ (green squares), and $5d$ (blue triangles) transition metals between the groups IV and XI. For $3d$ metals, their magnetic phases (partially transparent red circles) are also considered as well as their nonmagnetic counterpart; the antiferromagnetic phase for Cr and Mn, and the ferromagnetic phase for Fe, Co, and Ni.
\label{fig:all}
}
\end{figure*}

\section{Results}
\label{sec:results}

\subsection{SHE and OHE in a few selected materials}

We present detailed results of $\sigma_\mathrm{SH}$ and $\sigma_\mathrm{OH}$ for W, Pt, and Cu, which are commonly used metals in spin-orbitronics. Figure~\ref{fig:SHE_OHE} shows the results of $\sigma_\mathrm{SH}$ and $\sigma_\mathrm{OH}$ as a function of the Fermi energy $E_\mathrm{F}$, which is varied assuming that the effective single-particle potential is fixed to the value at the true Fermi energy $E_\mathrm{F}^\mathrm{true}$. Blue solid lines and orange dashed lines are the results obtained by considering the full expression for the velocity and by considering only the group-velocity-like contribution and ignoring the anomalous position contribution in Eq.~\eqref{eq:velocity_final}, respectively. In general, the anomalous position barely affects $\sigma_\mathrm{SH}$'s [Figs.~\ref{fig:SHE_OHE}(a-c)], but $\sigma_\mathrm{OH}$'s [Figs.~\ref{fig:SHE_OHE}(d-f)] are strongly affected by the anomalous position contribution. This is consistent with the model analysis presented in Sec.~\ref{subsec:orbital_origin_A}, which shows that the anomalous position is explicitly orbital-dependent but spin-independent [Eq.~\eqref{eq:anomalous_position_model}]. We remark that this is a property of basis states, and the negligible role of the anomalous position in SHE implies that the basis states are nearly diagonal in the spin space. In general, however, the basis states may depend on the spin depending on how the basis states are prepared. For example, in obtaining maximally localized Wannier functions, both orbital and spin characters of the basis states are mixed during the localization procedure if the SOC is present. Nonetheless, our numerical results imply that this effect is negligible for transition metals.

For W, the magnitude of $\sigma_\mathrm{OH}$ is reduced at $E_\mathrm{F}=E_\mathrm{F}^\mathrm{true}$ upon considering the anomalous position, but the magnitude increases at $E_\mathrm{F} - E_\mathrm{F}^\mathrm{true} \approx - 4\ \mathrm{eV}$ [Fig.~\ref{fig:SHE_OHE}(d)]. A similar trend is also observed for Pt [Fig.~\ref{fig:SHE_OHE}(e)]. In general, the results obtained by considering the full velocity expression reveal that the variation of $\sigma_\mathrm{OH}$ with respect to the Fermi energy is smaller as compared to the values obtained by considering only the group-velocity-like term. Strikingly, by including the anomalous position, the OHE in Pt is significantly suppressed at the true Fermi energy to $\sigma_\mathrm{OH}=144\ (\hbar/e)(\Omega\cdot\mathrm{cm})^{-1}$. This is in stark contrast to the previous works predicting large OHE in Pt: References~\cite{Jo2018, Salemi2022} predict a value of $\sigma_\mathrm{OH}\approx 3000\ (\hbar/e)(\Omega\cdot\mathrm{cm})^{-1}$. However, we would avoid putting strong emphasis on this prediction and argue that accurate estimation of the orbital Hall conductivity in Pt may be difficult due to the large slope of $\sigma_\mathrm{OH}$ near $E_\mathrm{F} \approx E_\mathrm{F}^\mathrm{true}$. The band structure and the Fermi energy depends on exchange correlation functionals, and excited states above the Fermi energy cannot be precisely computed within the DFT. In fact, experimental results of Ref.~\cite{Sala2022} suggest that the OHE in Pt may be substantial. Nonetheless, we argue that OHE of Pt in experiments could critically be affected by the chemical environment near Pt atoms, e.g. work function difference at the interface, oxidization and charge transfer, etc. because the result may significantly differ with slight modification of the band structure near the Fermi energy. Finally, in Cu, the sign of the OHE becomes negative upon including the anomalous position in the calculation, and the magnitude is appreciable, $\sigma_\mathrm{OH}=-788\ (\hbar/e)(\Omega\cdot\mathrm{cm})^{-1}$. We note that Refs.~\cite{Jo2018, Salemi2022} predict the positive sign of $\sigma_\mathrm{OH}$ in Cu, which is opposite to our prediction. Also, recent experiments have reported the positive sign of the OHE in Cu~\cite{Rothschild2022, Xu2023}, but the property of a Cu film may be change by surface oxidization~\cite{An2016, Ding2020, Kim2021, Ding2022, Ding2922unidirectional}, which leads to strongly enhanced orbital Rashba effect~\cite{Go2021orbital}.

Another important feature that is commonly found in both Pt and Cu is that as $E_\mathrm{F}$ increases beyond $E_\mathrm{F}^\mathrm{true}$, $\sigma_\mathrm{OH}$ becomes negative and eventually converges to zero as $E_\mathrm{F}$ further increases. We argue that this is a direct consequence of the fact that the electrons cannot carry OAM if the $d$ shell becomes fully occupied. We note that previous works \cite{Jo2018, Salemi2022} predicted an anomalous feature that $\sigma_\mathrm{OH}$ does not converge to zero even if $E_\mathrm{F}$ increases such that the $d$-shell is filled. We note that some experiments on Cu films have found evidence that pure Cu blocks the transmission of OAM~\cite{Kim2021, Ding2020}. We predict that for electron-doped Cu films, e.g. by surface deposition of alkali metals, the OHE must be negligibly small.

\subsection{Summary of the results for transition metals}

Figure~\ref{fig:all} shows the summary of the computed (a) $\sigma_\mathrm{SH}$ and (b) $\sigma_\mathrm{OH}$ of the transition metals between the group IV and XI. Their numerical values at the Fermi energy are listed in Table~\ref{tab:SHE_OHE_dat}. We find that the overall trend of $\sigma_\mathrm{SH}$ is similar to what has been found in previous works~\cite{Tanaka2008, Jo2018, Salemi2022}. Generally $\sigma_\mathrm{SH}$ is negative when the $d$-shell is approximately less than half-filled (groups IV--VII) and becomes positive when the $d$-shell becomes more occupied (groups VIII--XI). For $3d$ metals, we present the results for both nonmagnetic and magnetic phases; For Cr and Mn, we consider antiferromagnetism, and for Fe, Co, and Ni, we consider ferromagnetism. For antiferromagnetic Cr and Mn, the values of $\sigma_\mathrm{SH}$'s are smaller than those computed for the nonmagnetic phase. Among the ferromagnetic elements, Co exhibits the most drastic change of $\sigma_\mathrm{SH}$ due to ferromagnetism: The nonmagnetic Co shows large positive value of $\sigma_\mathrm{SH}= 803\ (\hbar/e)(\Omega\cdot\mathrm{cm})^{-1}$, but the ferromagnetic Co shows an order of magnitude smaller magnitude and negative sign of the spin Hall conductivity $\sigma_\mathrm{SH}= -87\ (\hbar/e)(\Omega\cdot\mathrm{cm})^{-1}$. On the other hand, Fe and Ni show only a slight change of $\sigma_\mathrm{SH}$ with ferromagnetism.

For the OHE in transition metals [Fig.~\ref{fig:all}(b)], our results show that the variation of the magnitude of $\sigma_\mathrm{OH}$ for the elements between the group IV and IX is generally small, with values between $4000\ (\hbar/e)(\Omega\cdot\mathrm{cm})^{-1}$ and $6000\ (\hbar/e)(\Omega\cdot\mathrm{cm})^{-1}$. On the other hand, Refs.~\cite{Jo2018,Salemi2022} found larger variation of the magnitude among the transition metals across the group IV--IX, and their maximum values are generally larger. The elements in the group X and XI exhibit  much smaller values of $\sigma_\mathrm{OH}$ compared to the elements between groups IV-IX. Interestingly, we find that the sign of OHE for some elements in the group X and XI negative, which is the case for Ni (nonmagnetic), Cu, Pd, Ag, and Au. We note that the magnitude of the orbital Hall conductivity is not negligible for Ni (nonmagnetic) and Pd, $\sigma_\mathrm{OH} \approx -1500\ (\hbar/e)(\Omega\cdot\mathrm{cm})^{-1}$, which may be experimentally confirmed. We remark that the previous works,  Refs.~\cite{Tanaka2008, Jo2018, Salemi2022}, predicted the positive sign of OHE for all transition metals, which is different from our result. The negative sign of the intrinsic OHE has been reported for Si, Ge, and $\alpha$-Sn~\cite{Baek2021}. Finally, for $3d$ elements, magnetism tends to make $\sigma_\mathrm{OH}$ smaller compared to the value computed for the nonmagnetic counterpart. This effect is more dramatic for ferromagnetic metals (Fe, Co, Ni), while for antiferromagnetic metals (Cr and Mn), the magnetism suppresses $\sigma_\mathrm{OH}$ only slightly.

\section{Discussion}
\label{sec:discussion}

We note that the mechanism of OHE due to the OAM-dependent anomalous position was pointed out by Jung \emph{et al.}~\cite{Jung2014}. However, the origin of the anomalous position was not clarified, and the form of the anomalous position was assumed from the analogy to the orbital Rashba effect [Eq.~\eqref{eq:orbital_Rashba_model}]. In our work, we have explicitly shown the orbital-origin of the anomalous position in the formal derivation in Sec.~\ref{sec:background}. 

A similar expression is also found in the mechanism of OHE proposed in Ref.~\cite{Go2018} by some of us. We remark that this mechanism, which relies on orbital hybridizations of energy bands, is different from what we have presented in Secs.~\ref{subsec:orbital_origin_A}. In Ref.~\cite{Go2018}, it is assumed that the position operator is simply given by $\hat{\mathbf{r}}=i\boldsymbol{\nabla}_\mathbf{k}$. Nonetheless, due to orbital hybridizations, $\mathbf{k}\times\mathbf{L}$ type anomalous position emerges in the description of an effective description in the Hamiltonian gauge. That is, Ref.~\cite{Go2018} has explicitly shown that 
\begin{eqnarray}
iU^\dagger (\mathbf{k})
[
\boldsymbol{\nabla}_\mathbf{k} U (\mathbf{k})
]
=
\lambda \mathbf{k}\times\mathbf{L}^\mathrm{(H)},
\label{eq:k_cross_L_2}
\end{eqnarray}
where $\lambda$ depends on the orbital hybridizations contained in the Hamiltonian. We remark that the anomalous position we discuss in this article is due to spatial overlap of the basis states and independent from the microscopic Hamiltonian, such as the relativistic correction by the electron-positron hybridization in the Dirac equation~\cite{Foldy1950, Vignale2010, Bi2013, Ado2023} or the Yafet term induced by the band hybridizations~\cite{Nozieres1973, Winkler2003, Engel2007}.

We have shown that our results on the OHE in transition metals exhibit both qualitatively and quantitatively different features compared to the previous works, Refs.~\cite{Tanaka2008, Jo2018, Salemi2022}. Reference~\cite{Salemi2022} employs the real-space representation of the basis states in first-principle methods, and $\hat{\mathbf{p}}/m$ is used as the velocity operator, where $\hat{\mathbf{p}}=-i\hbar\boldsymbol{\nabla}$ is the canonical momentum operator and $m$ is the electron's rest mass. It also considers semicore states, which give rise to a finite value even when the Fermi energy is below the bottom of the valence bands. On the other hand, the current manuscript considers only the valance states with Wannier functions. In Ref.~\cite{Jo2018}, the anomalous position is entirely neglected because the information of the dipole matrix elements [Eq.~\eqref{eq:dr_matrix}] is missing in the tight-binding description. Similarly, other theoretical works that employed tight-binding models did not consider the anomalous position~\cite{Bernevig2005, Tanaka2008, Kontani2009, Go2018, Canonico2020a, Canonico2020b, Cysne2021, Bhowal2021, Cysne2022}. We argue that for consistent description of orbital response phenomena, microscopic information of the basis states are crucial because they strongly affect the dipole matrix elements of the position. This implies that first-principles methods are probably the most suitable way to quantitatively predict orbital response phenomena.

Wang \emph{et al.} have investigated the anomalous Hall effect by Wannier interpolation and explicitly compared the contribution due to the anomalous position to the other contributions~\cite{Wang2006}. They have shown that the anomalous position contribution is negligible in anomalous Hall effect because the commutator with the Hamiltonian [the second line in Eq.~\eqref{eq:velocity_final}] cancels the energy difference appearing in the denominator of the Kubo formula [Eq.~\eqref{eq:Kubo}]. However, we emphasize that the cancellation does not apply to the orbital current. For the matrix elements of the velocity operator, the main reason for the cancellation of the energy gap in the denominator relies on the fact that the velocity operator is written as the commutator between the position operator and the Hamiltonian~[Eqs.~\eqref{eq:position_final} and \eqref{eq:velocity_final}]. The orbital current in Eq.~\eqref{eq:conventional_orbital_current} is defined as the product of the velocity operator and OAM operator, and thus unless the OAM operator is diagonal in the Hamiltonian gauge, the transitions induced by the OAM operator prevent the anomalous position term producing the exact energy difference in the denominator of the Kubo formula. We remark that the same problem persists for spin current, but since the anomalous position is mainly orbital-dependent [Eq.~\eqref{eq:anomalous_position_model}], the anomalous position and the spin operator commute with  each other. 

Finally, we note that our computed results of $\sigma_\mathrm{OH}$ cannot be directly compared to the orbital accumulation at a surface or the torque on local magnetic moments, which requires consistent treatment of the OAM non-conserving interactions, e.g. by the continuity equation~\cite{Haney2010, Go2020b}. Thus, the results shown in Fig.~\ref{fig:all} and Tab.~\ref{tab:SHE_OHE_dat} should be taken with caution. Because the OAM is not conserved by the crystal field potential, which persists regardless of SOC, a part of the orbital current may be absorbed by the crystal. This in general leads to the suppression of the effective orbital Hall conductivity~\cite{Choi2023}. On spin current, Shi \emph{et al.} proposed to use the definition of spin current as the total time-derivative of the spin dipole as it has a direct relation to the spin accumulation at a surface. We note that the cancellation issue discussed in the above paragraph is solved by adopting the definition of Ref.~\cite{Shi2006}. Recently, Liu \emph{et al.} proposed that the spin Hall conductivity tensor becomes concisely expressed in terms of geometric quantities of Bloch states by using this definition of spin current~\cite{Liu2023}. We leave the investigation of an alternative definition of orbital current and its first-principles calculation in real materials for future work. However, we still remark that because the anomalous position has orbital dependence, in general, consistent treatment of the position operator is going to still be important in any $\mathbf{k}$-space definition of the orbital current.

\section{Conclusion}
\label{sec:conclusion}

In this work, we have shown that the anomalous position naturally appears on top of the canonical position in $\mathbf{k}$-space as a gauge correction to the position operator. The microscopic origin of the anomalous position is the dipole matrix element between basis functions in real space, and thus it is explicitly orbital-dependent. From the analysis of a simple model, we have demonstrated that it results in the non-Abelian Berry curvature, which is directly proportional to the OAM operator near $\mathbf{k}=0$. Thus, the anomalous position alone can generate OHE, and this mechanism is independent from the previously proposed mechanism due to the orbital hybridization~\cite{Go2018}. By first-principles calculation of the transition metals between the group IV--XI, we have found that the anomalous position barely affects SHE. However, we find the crucial role of the anomalous position in OHE. Our results are different from the previous theoretical works on the OHE in transition metals~\cite{Tanaka2008, Jo2018, Salemi2022}, which would require a comparative study on different methods. In particular, we predict the negative sign of OHE for some elements in the groups X and IX, such as nonmagnetic Ni, Cu, Pd, Ag, and Au. Because the magnitude of OHE in these elements is far from being negligible, the negative sign of OHE may be experimentally confirmed.

\begin{table*}
\begin{center}
\centering
\begin{tabular}{M{2.0cm} | M{3.5cm} M{2.1cm} M{2.1cm} M{1.5cm} M{2.2cm} M{3.5cm}}
\hline \hline
Material & lattice constant $(a_0)$ & $R_\mathrm{MT}\ (a_0)$ & $K_\mathrm{max}\ (a_0^{-1})$ & $l_\mathrm{max}$ & DFT $\mathbf{k}$-mesh & Interpolation $\mathbf{k}$-mesh \\
 \hline
hcp Ti & $a=5.54, \ c=8.80$ & $2.65$ & $4.5$ & $12$ & $16\times 16\times 12$ & $256\times 256\times 192$ \\ 
bcc V & $a=5.73$ & $2.42$ & $4.5$ & $12$ & $16\times 16\times 16$ & $256\times 256\times 256$\\ 
bcc Cr & $a=5.50$ & $2.32$ & $4.5$ & $12$ & $16\times 16\times 16$  & $256\times 256\times 256$\\ 
fcc Mn & $a=6.6.3$ & $2.29$ & $4.5$ & $12$ & $16\times 16\times 16$  & $256\times 256\times 256$\\ 
bcc Fe & $a=5.42$ & $2.29$ & $4.5$ & $12$ & $16\times 16\times 16$  & $256\times 256\times 256$\\ 
hcp Co & $a=4.74, \ c=7.69$ & $2.65$ & $4.5$ & $12$ & $16\times 16\times 12$ & $256\times 256\times 192$\\ 
fcc Ni & $a=6.65$ & $2.30$ & $4.5$ & $12$ & $16\times 16\times 16$  & $256\times 256\times 256$\\ 
fcc Cu & $a=6.83$ & $2.29$ & $5.0$ & $12$ & $16\times 16\times 16$  & $256\times 256\times 256$\\
\hline
hcp Zr & $a=6.10, \ c=9.73$ & $2.80$ & $4.5$ & $12$ & $16\times 16\times 12$ & $256\times 256\times 192$\\ 
bcc Nb & $a=6.28$ & $2.65$ & $4.5$ & $12$ & $16\times 16\times 16$  & $256\times 256\times 256$\\ 
bcc Mo & $a=5.95$ & $2.52$ & $5.0$ & $12$ & $16\times 16\times 16$  & $256\times 256\times 256$\\ 
hcp Tc & $a=5.18, \ c=8.30$ & $2.49$ & $5.0$ & $12$ & $16\times 16\times 12$ & $256\times 256\times 192$\\ 
hcp Ru & $a=5.15, \ c=8.13$ & $2.46$ & $4.5$ & $12$ & $16\times 16\times 12$ & $256\times 256\times 192$\\ 
fcc Rh & $a=7.18$ & $2.47$ & $5.0$ & $12$ & $16\times 16\times 16$  & $256\times 256\times 256$\\
fcc Pd & $a=7.35$ & $2.53$ & $5.0$ & $12$ & $16\times 16\times 16$  & $256\times 256\times 256$\\
fcc Ag & $a=7.73$ & $2.66$ & $5.0$ & $12$ & $16\times 16\times 16$  & $256\times 256\times 256$\\
\hline
hcp Hf & $a=6.05, \ c=9.54$ & $2.80$ & $5.0$ & $12$ & $16\times 16\times 12$ & $256\times 256\times 192$\\ 
bcc Ta & $a=6.24$ & $2.63$ & $5.0$ & $12$ & $16\times 16\times 16$  & $256\times 256\times 256$\\ 
bcc W & $a=5.96$ & $2.52$ & $4.5$ & $12$ & $16\times 16\times 16$  & $256\times 256\times 256$\\ 
hcp Re & $a=5.22, \ c=8.43$ & $2.52$ & $4.5$ & $12$ & $16\times 16\times 12$ & $256\times 256\times 192$\\ 
hcp Os & $a=5.16, \ c=8.43$ & $2.52$ & $5.0$ & $12$ & $16\times 16\times 12$ & $256\times 256\times 192$\\ 
fcc Ir & $a=7.26$ & $2.50$ & $5.0$ & $12$ & $16\times 16\times 16$   & $256\times 256\times 256$\\
fcc Pt & $a=7.42$ & $2.56$ & $4.5$ & $12$ & $16\times 16\times 16$   & $256\times 256\times 256$\\
fcc Au & $a=7.71$ & $2.65$ & $5.0$ & $12$ & $16\times 16\times 16$   & $256\times 256\times 256$\\
\hline \hline
\end{tabular}
\caption{
\label{tab:params}
Parameters used for the DFT calculation in the FLAPW method. For the body-centered cubic (bcc) and face-centered cubic (fcc) structures, the lattice constant in the cubic cell convention $a$ is shown. For the hexagonal close packed (hcp) structure, both in-plane (a) and out-of-plane (c) lattice constants are shown. $a_0$ is the Bohr radius. $R_\mathrm{MT}$ is the muffin-tin radius, $K_\mathrm{max}$ is the plane wave cutoff, $l_\mathrm{max}$ is the maximum number of the harmonic expansion in the muffin-tin. The DFT $\mathbf{k}$-mesh is used for the self-consistent calculation for converging the electronic states, and the interpolation $\mathbf{k}$-mesh is used for the computation of the response function [Eq.~\eqref{eq:Kubo}] in the Wannier representation.
}
\end{center}
\end{table*}

\begin{table}[t!]
\begin{center}
\centering
\begin{tabular}{M{2.5cm} | M{2.5cm} M{2.5cm}}
\hline \hline
Material & $\sigma_\mathrm{SH}$ & $\sigma_\mathrm{OH}$ \\
\hline
hcp Ti & $-17$ & $4304$ \\
bcc V & $-43$ & $4492$ \\
bcc Cr (N) & $-162$ & $5829$ \\
bcc Cr (AF) & $-68$ & $4799$ \\
fcc Mn (N) & $-188$ & $6087$ \\
fcc Mn (AF) & $-37$ & $5066$ \\
bcc Fe (N) & $456$ & $6305$ \\ 
bcc Fe (F) & $587$ & $2345$ \\ 
hcp Co (N) & $803$ & $5762$ \\
hcp Co (F) & $-87$ & $2155$ \\
fcc Ni (N) & $1548$ & $-1438$ \\
fcc Ni (F) & $2013$ & $886$ \\
fcc Cu & $83$ & $-778$ \\
\hline
hcp Zr & $-30$ & $4460$ \\
bcc Nb & $-74$ & $4018$ \\
bcc Mo & $-254$ & $3825$ \\
hcp Tc & $-72$ & $3731$ \\
hcp Ru & $135$ & $5545$ \\
fcc Rh & $987$ & $4982$ \\
fcc Pd & $1111$ & $-1870$ \\ 
fcc Ag & $55$ & $-856$ \\
\hline 
hcp Hf & $50$ & $4310$ \\
bcc Ta & $-160$ & $3801$ \\
bcc W & $-788$ & $4664$ \\
hcp Re & $-456$ & $4316$ \\
hcp Os & $-40$ & $4814$ \\
fcc Ir & $321$ & $4434$ \\
fcc Pt & $2212$ & $144$ \\
fcc Au & $359$ & $-1020$ \\
\hline \hline
\end{tabular}
\caption{
\label{tab:SHE_OHE_dat}
Spin Hall conductivities ($\sigma_\mathrm{SH}$) and orbital Hall conductivities ($\sigma_\mathrm{OH}$) of the transition metals between the groups IV and XI at the Fermi energy, which are written in the unit of $(\hbar/e)(\Omega\cdot\mathrm{cm})^{-1}$. For $3d$ transition metals, N, AF, and F in the parentheses indicate normal, antiferromagnetic, and ferromagnetic phases, respectively.}
\end{center}
\end{table}

\begin{acknowledgements}
D.G. and Y.M. acknowledge discussion with Daegeun Jo, Peter Schmitz, and Frank Freimuth and thank Cong Xiao for insightful comments on the manuscript. D.G. and Y.M. gratefully acknowledge the J\"ulich Supercomputing Centre for providing computational resources under project jiff40. D.G. and Y.M. acknowledge the funding by the Deutsche Forschungsgemeinschaft (DFG, German Research Foundation) $-$ TRR 173/2 $-$ 268565370 (project A11), TRR 288 $-$ 422213477 (project B06). H.W.L was supported by the Samsung Science and Technology Foundation (BA-1501-51). P.M.O. acknowledges the funding by the Wallenberg Foundation (Grant no. 2022.0079) and the Wallenberg Initiative Materials Science for Sustainability (WISE). 
\end{acknowledgements}

 \clearpage
\bibliography{bib_OHE}
\end{document}